\RequirePackage{fix-cm}
\documentclass[pdftex, epjc3,final]{svjour3}      
\RequirePackage[T1]{fontenc}
\smartqed  % flush right qed marks, e.g. at end of proof
\RequirePackage{graphicx}
\RequirePackage{mathptmx}      % use Times fonts if available on your TeX system
\RequirePackage{flushend}
\RequirePackage{booktabs}
\RequirePackage[numbers,sort&compress]{natbib}
\RequirePackage{amsmath, amsfonts, amssymb, mathrsfs, mathtools, xfrac, physics, bm, amsbsy, marvosym, gensymb}
\RequirePackage{bigstrut}
\RequirePackage{float}
\RequirePackage[labelfont=bf, font=small]{caption}
\RequirePackage{subcaption}
\RequirePackage[colorlinks,citecolor=blue,urlcolor=blue,linkcolor=blue]{hyperref}
\journalname{Eur. Phys. J. C}

\providecommand{\abs}[1]{\lvert #1 \rvert}

\expandafter\let\expandafter\hbar\csname ?-\string\hbar\endcsname

\def\al{\alpha}
\def\b{\beta}

\def\th{\theta}
\def\lam{\lambda}

\def\D{\Delta}
\def\de{\delta}

\def\chic#1{{\scriptscriptstyle #1}}
\def\xy{\rm{\chic{12}}}
\def\xz{\rm{\chic{13}}}
\def\be{\begin{equation}}
\def\ee{\end{equation}}
\def\bea{\begin{eqnarray}} 
\def\eea{\end{eqnarray}} 
\def\ba{\begin{array}}
\def\ea{\end{array}}
\def\ni{\noindent}
\def\hs{\hspace}                                                                    

\def\hsp{\hspace{.03cm}}
\def\hsn{\hspace{-0.03cm}}   
\def\ben{\begin{enumerate}}                                                                                                            
\def\een{\end{enumerate}}
\def\bei{\begin{itemize}}
\def\eni{\end{itemize}}

\def\mc{\mathcal}

\def\mtt{\mathtt}

\def\Oxy{{\cal O}_{12}}
\def\Oxz{{\cal O}_{13}}
\def\Oyz{{\cal O}_{23}}
\def\G{\rm\Gamma}

\def\lan{\langle}
\def\ran{\rangle}

\begin{document}

\title{Oscillation tomografy study of Earth's composition and density with atmospheric neutrinos}
\subtitle{}

\author{Juan Carlos D'Olivo\thanksref{e1,addr1} \and Jos\'e Arnulfo Herrera Lara\thanksref{e2,addr1} \and Ismael Romero\thanksref{e3,addr2} \and Oscar A. Sampayo\thanksref{e4,addr2}}

\thankstext{e1}{e-mail: dolivo@nucleares.unam.mx}
\thankstext{e2}{e-mail: fis\_pp@ciencias.unam.mx}
\thankstext{e3}{e-mail: ismaelromero\_@hotmail.com}
\thankstext{e4}{e-mail: sampayo@mdp.edu.ar}

\institute{Instituto de Ciencias Nucleares, Universidad Nacional Aut\'onoma de M\'exico, 
Circuito Exterior, Ciudad Universitaria, 04510 CDMX, M\'exico.\label{addr1}
          \and Instituto de F\'isica de Mar del Plata (IFIMAR)\\
CONICET, UNMDP\\ Departamento de F\'isica,
Universidad Nacional de Mar del Plata \\
Funes 3350, (7600) Mar del Plata, Argentina.\label{addr2}
}

\date{Received: date / Accepted: date}
% The correct dates will be entered by the editor

\maketitle

\begin{abstract}
Knowledge of the composition of the Earth's interior is highly relevant to many geophysical and geochemical problems. Neutrino oscillations are modified in a non-trivial way by the matter effects and can provide valuable and unique information not only on the density but also on the chemical and isotopic composition of the deep regions of the planet. In this paper, we re-examine the possibility of performing an oscillation tomography of the Earth with atmospheric neutrinos and antineutrinos to obtain information on the composition and density of the outer core and the mantle, complementary to that obtained by geophysical methods.  Particular attention is paid to the D$^{\prime \prime}$ layer just above the core-mantle boundary and to the water (hydrogen) content in the mantle transition zone. Our analysis is based on a Monte-Carlo simulation of the energy and azimuthal angle distribution of $\mu$-like events generated  by neutrinos. Taking as reference a model of the Earth consisting of 55 concentric layers with constant densities determined from the PREM, we evaluate the effect on the number of events due to changes in the composition and density of the outer core and the mantle. To examine the capacity of a detector like ORCA to resolve such variations, we construct regions in planes of two of these quantities where the statistical significance of the discrepancies between the reference and the modified Earth are less than $1\sigma$. The variations are implemented in such a way that the constraint imposed by both the total mass of the Earth and its moment of inertia are verified. 
\end{abstract}

\section{Introduction}
\label{intro}

Aside from its intrinsic interest, a detailed description of the inner parts of the Earth is essential for a proper understanding of basic geological phenomena such as volcanology, earthquakes, plate tectonics, and mountain building
\cite{tarbuck2009earth,fowler:2005}. According to current knowledge, the Earth's internal structure is stratified and consist of successive layers with different chemical, geological, and physical properties. Most of the information about the layers and the boundaries between them has been acquired by examining how seismic waves created by natural earthquakes are refracted and reflected as they propagate through the Earth \cite{Lay:1995,Aki:2002}. Valuable information has also been obtained from measurements of magnetic and gravitational fields, observations of the planet's moment of inertia and precessional motion, and physical, chemical, and mineralogical analyzes of meteorites and xenoliths. Based on chemical composition, three main layers have been identified within our planet: the crust, the mantle, and the core. The mantle and the core are further subdivided into two regions each.

The mantle surrounds the core and extends from beneath the thin crust to a depth of 2,900 km. It consists mainly of dense silicate rocks rich in iron and magnesium. Seismic wave velocities in the mantle show three discontinuities at depths of 410, 660 and 2,700 km \cite{Helffrich:2001}. The first two correspond to the edges of the transition zone between the upper and lower mantle, and are best explained by mineral phase transformations without compositional changes. The third discontinuity is typically a 2.5-3.0\% increase for both S- and P-waves observed at the top of the D$^{\prime\prime}$ layer, a transition shell about 200-300 km thick, at the base of the mantle, which presents a variety of seismic anomalies and is presumably the source of the large mantle plumes \cite{Loper:1995,Laywg:1998}. This layer is not yet fully understood, many of its characteristic may be attributed to the discovered MgSi${\rm O}_3$ postperovskite phase \cite{Murakami:2004}, but compositional differences may also play an important role in addition to the phase transformation \cite{Hiroshe:2021}. Regarding the core, it is well established that it is composed primarily of an iron-nickel alloy, with Ni/Fe\,$\sim 0.06$. The inner core is solid, while the lower density and absence of S-wave propagation are indications of a liquid outer core. The density deficit in the outer core cannot be simply explained by a difference in the state, but it requires about 5-10 wt\% (weight percent) of lower atomic weight elements to reduce its density and melting point \cite{Birch:1964, Poirier:1994}. 

Good estimates of the abundance and distribution of ``light''  elements in the core are essential to understanding the formation and evolution of the Earth, as well as how the core and mantle interact in the region around the core-mantle boundary (CMB) \cite{Litasov:2016,Hiroshe:2021}. The various processes that occur in this interface are highly influenced by heterogeneous structures at or near it. The  great disparity in density prevents direct convection movement through the CMB, and nearby regions control the transfer of heat and material \cite{Wookey:2008}. These processes strongly affect the convection in the mantle, responsible for the plate motion and continental drift, and the more vigorous convective flow in the outer core that is believed to be at the origin of the Earth's magnetic field (i.e., the functioning of the geodynamo) \cite{Lister:1995,Roberts:2013}. Due to vigorous convection, the liquid part of the core is usually assumed to have a uniform composition. However, seismological evidence indicates the possible existence of a $\sim 300$ km thick region on top of the outer core (E$^\prime$ layer), which shows anomalous low seismic velocities. This region is likely to be less dense than the rest of the outer core, but simply increasing the concentration of light elements also produces higher speeds, in contradiction with observations. None of the mechanisms that have been considered are without complications \cite{Brodholt:2017} and further observations are needed before a satisfactory explanation of the origin and nature of the E' layer can be formulated. 

The study of seismic wave propagation and normal mode oscillation is undoubtedly the most effective and reliable method to search the Earth's deep structures and process. Nevertheless, the impressive progress done has not been accompanied with a concomitant improvement in the precision of the density estimations \cite{Bolt:1995}. They are done through an underdetermined inversion problem, performed in two steps: the spatial distribution of seismic wave velocities is first inferred from seismological data and then the density distribution is inferred from the seismic velocities using some empirical relation \cite{Geller:2003}.  Such procedure allows the average density along a path to be estimated with an uncertainties of about 5\% \cite{Bolt:1995} for the mantle and presumably larger for the core. For example, the density jump at the inner-core boundary, which play an important role in the maintenance of the geodynamo, has been inferred to be of 0.82 with an error of more than 20\% \cite{Master:2003}. While the density distribution can be obtained from seismological remote sensing, the compositional structure of the Earth  \cite{McDonough:1995,allegre1995chemical} has been much more difficult to determine. Thus, the compositions of the lower mantle and the core remain quite uncertain, despite significant advances in recent years.  Since in situ sampling is impossible, estimations are done by comparing density and sound velocity data from seismological observations with those from laboratory experiments and theoretical calculations \cite{Zhang:2016}. Due to technical limitations, it has been difficult to perform reliable experiments for molten samples under high pressure and temperature conditions and the available information is still insufficient to infer the core composition. The most likely light elements in the core are oxygen, silicon, carbon, nitrogen, sulfur, and hydrogen \cite{Li:2021,Hiroshe:2021}, but there is still no consensus on the nature and proportion of the components. 

In addition to the metallic core, significant amounts of hydrogen can be incorporated within the mantle, in silicate minerals, melts, and hydrous fluids. This is done in a variety of chemical species (OH, H$_2$, $\cdots$) generically referred to as ``water" \cite{Hirschmann:2006,Peslier:2017,Othani:2020}. The abundance and distribution of water has influenced not only the evolution, dynamics state and thermal structure of the deep interior, but also the evolution of the crust and hydrosphere \cite{Bodnar:2013}.
Even small amount of water can affect properties like melting  temperature, rheological properties, electrical conductivity, and seismic velocities of the mantle. How water is transported into Earth's deep interior and how it is distributed are today open questions. Almost no constraints exists on the water content of the lower mantle. The existing laboratory data cannot be used to infer the water content from geophysical observations at these high pressure and temperature conditions. Low-velocity regions have been observed near the top of the lower mantle and directly above the CMB \cite{Peslier:2017}. If the low-velocity is interpreted as caused by partial melting, then some water is likely be present in these regions \cite{Garnero:2016}. Water could instead be stored in the lowest parts of the mantle in Al-postperovskite \cite{Townsend:2016}. There is a wide consensus that the mantle transition zone (MTZ), at 410-660 km deep, is a potential reservoir of water because its main mineral constituents can store up to $\sim 3\%$ wt water \cite{Bodnar:2013}. However, the amount of water contained there is poorly constrained. Some water-rich inclusions recently found in diamonds \cite{Pearson:2014,tschauner2018ice} suggest a wet MTZ, but it is not clear if they  are representative of the typical water content of the deep mantle or reflect local conditions. Geophysical methods (electrical conductivity and seismic observations) provides constraints on the water distribution in a global scale \cite{karato2011water}. The inferences with these approaches are not direct and, despite the effort involved, a wide range of values have been reported in the literature \cite{karato:2020,Fei:2017}. As noticed in Ref. \cite{karato:2020}, geophysical estimates have large uncertainties and the water content in the MTZ can be heterogeneous, having significant lateral variability as revealed by a recent novel approach \cite{munch:2020}.

From the comments in the previous paragraph it is apparent how useful it would be to have additional experimental techniques, unaffected by the same uncertainties, which could provide complementary and independent information about the deep interior of the Earth. A promising candidate is neutrino tomography \cite{Winter:2006vg}.  The basic idea is that neutrino propagation within the Earth is affected by their interactions with the particles present in the terrestrial matter. The cross section for neutrino interactions increase with energy and, in the case of absorption tomography, the density profile can be reconstructed from the attenuation of the flux of very high energy ($\gtrsim 10$ TeV) neutrinos passing through the Earth \cite{Volkova:1974xa, Wilson:1983an, Ralston:1999fz, Jain:1999kp, GonzalezGarcia:2007gg, Reynoso:2004dt, Romero:2011zzb,Donini:2018tsg}. Another option is oscillation tomography, which takes advantage of the matter effect on flavor oscillations \cite{Wolfenstein:1977ue, Barger:1980tf, Mikheev:1986gs} of lower energy (MeV to GeV) neutrinos. In a medium, the transition probabilities between active neutrinos depend on the number density of electrons $n_e$ along the neutrino trajectory, which is proportional to the product of the matter density $\rho$ times the average ratio $Z/A$ of the atomic number $Z$ and the mass number $A$ \cite{Nicolaidis:1987fe, Nicolaidis:1990jm, Winter:2006vg, Borriello:2009ad, Winter:2015zwx, Rott:2015kwa, VanElewyck:2017dga, Bourret:2017tkw, DOlivo:2020ssf, Kelly:2021jfs, Denton:2021rgt, Upadhyay:2021kzf}. 

In this paper, we re-examine the feasibility of studying the internal structure of the Earth using atmospheric neutrino oscillation tomography.  We analyze the ability of a detector such as ORCA to resolve deviations both of density and composition with respect to a standard Earth modeled in terms of 55 concentric shells integrated in five main layers, corresponding to the inner and outer core, the lower and upper mantle, and the crust. Constant densities are assigned to the shells from the mean value of the PREM densities within each shell.  In our scheme, unlike other work on the subject, the densities of the main layers can be modified in a manner consistent with the well-measured total mass and moment of inertia of the Earth. In contrast, $Z/A$ is a function of the chemical and isotopic composition of the medium and is not subject to either of these constraints. Therefore, in principle, one could constrain the allowed values of the density and composition of the Earth's deeper regions  by studying the effects that changes in these quantities have on the events produced by atmospheric neutrinos after traversing the Earth, holding the total mass and moment of inertia of the planet fixed.  We focus on the possible application to obtain information on the composition and density of the lower mantle regions above the CMB, in particular, the D$^{\prime \prime}$ region, paying special attention to the content of light elements, more specifically hydrogen.  With the exception of hydrogen ($(Z/A)_H = 1$), all other light elements have an almost equal number of protons and neutrons and hence $Z/A\cong 0.5$  for them. Thus, the presence of a significant amount of hydrogen (water) would produce appreciable $Z/A$ changes compared to those in dry regions. In this sense, information from neutrino oscillation tomography could give valuable information on the water content in the lower mantle and core. We also allow for some variation in the location of the boundary between the lower and upper mantle as an effective way to account for the transition region between these two layers. 

The paper is organized as follows. In Sect. \ref{sec:structure} we present a model of the Earth's structure with 55  shells integrated within 5 main layers. In Sect. \ref{sec:osc} we briefly review the formalism of matter neutrino oscillations and describe the algorithm to calculate the transition  probabilities. In Sect. \ref{sec:simu} we determine the number of $\mu$-like neutrino events in a detector
such as ORCA and the effects that changes in the composition  and density of the outer core and lower mantle have on this observable. The results and final comments are presented in Sect.\ref{sec:results}, where we carry out a Monte Carlo simulation of the number of  $\mu$-like events and apply it to test different composition models of the outer core and mantle, changing also the density.

%%%%%%%%%%%%%%%%%%%%%%%%%%%%%%%%%%%%%%%%%%%%%%%%%%%%%%%%%%%%%%%%%
\section{Model of the Earth's structure}
\label{sec:structure}
The observed lateral variations in the Earth's properties are much less pronounced than the vertical variations. Therefore, the internal structure of the Earth can be well approximated by one-dimensional spherically symmetric models of seismic velocities, attenuation, and density as a function of depth \cite{dziewonski1981preliminary,Kennett:1995}.The most widely used of these models for seismic tomography has been the Preliminary Reference Earth Model (PREM) \cite{dziewonski1981preliminary}. This model represents the mean properties of the Earth as a function of the radial distance $r$ and was designed to fit different data sets and some basic data of the planet (radius, mass, and moment of inertia). In this study,  we use a spherical model consisting of 55 concentric shells, each with a constant density equal to the average value of the PREM densities in the shell. The set of shells is divided into five large layers demarcated by concentric spheres of different radii: inner core ($IC$), outer core ($OC$), lower mantle ($M_1$), upper mantle ($M_2$), and crust ($C$). In Table \ref{tab:compo} we give the values for the standard composition and the corresponding radii.
\begin{figure}[h!]
\centering
\includegraphics[width=0.8\textwidth]{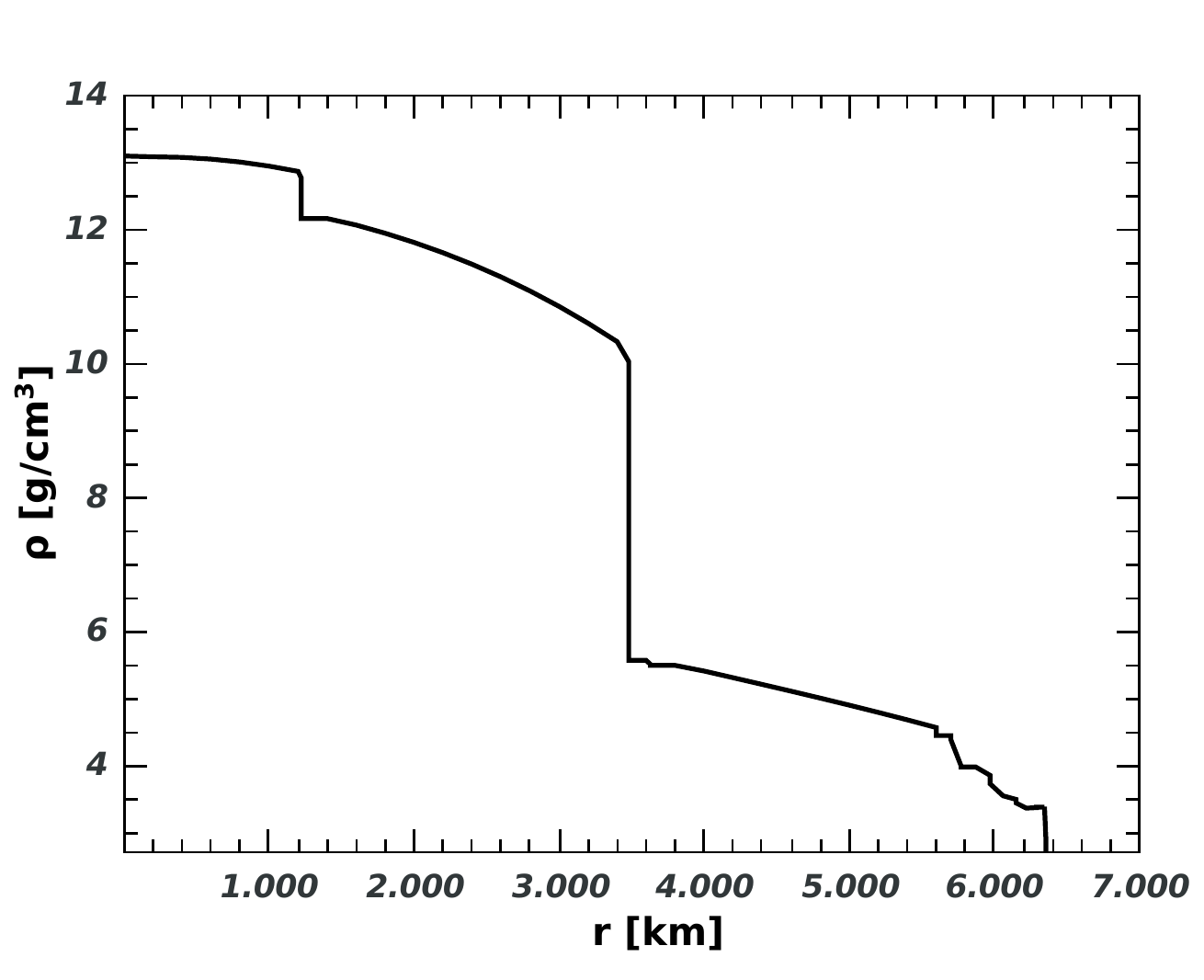}	
\caption{PREM density profile of the Earth with 55 shells.}
\label{fig:prem55}
\end{figure}

\begin{table}[h]
\centering
\begin{tabular}{@{}|c|c|c|c|@{}}
\toprule
\textbf{Layer} & \textbf{$n^0$ of Shells} & \textbf{$R_{inf}$ - $R_{sup}$ {[}km{]}} & \textbf{Z/A} \\ \midrule
Inner Core   & 7              & 0 - 1221.5         & 0.4691 \\
Outer Core   & 13             & 1221.5 - 3480      & 0.4691 \\
Lower Mantle & $N_{M_1}$-$21$ & $3480$ - $R_{M_1}$ & 0.4954 \\
Upper Mantle & $49$-$N_{M_1}$ & $R_{M_1}$ - $6346$ & 0.4954 \\
Crust        & 6346-6371      & 6346 - 6371        & 0.4956 \\ \bottomrule
\end{tabular}
\caption{Compositions of the main Earth layers.}
\label{tab:compo}
\end{table}

The primary information on the Earth's density as a function of $r$ comes from the total mass of the Earth 
$
M_{_{\hsn\oplus}} = 5.9724 \times 10^{27} {\rm g}
$
and its mean moment of inertia about the polar axis 
$
I_{_{\hsn\oplus}} = 8.025 \times 10^{44} {\rm g} \hs{0.03 cm}{\rm cm}^2
$
\cite{Williams:1994}. 
From these two quantities and $R_{_{\hsn\oplus}} = 6371$ km for the Earth's radius, one gets  $I_{_{\hsn\oplus}} \simeq 0.33 M_{_{\hsn\oplus}} R^2_{_{\hsn\oplus}}$ that is noticeably smaller than the moment of inertia of a homogeneous sphere of the same radius ($0.40 M_{_{\hsn\oplus}} R^2_{_{\hsn\oplus}}$). This corroborates that there must be a concentration of mass towards the center of the planet or, in other words, that the inner regions are denser than average. We require that our model satisfy both constraints:
\begin{eqnarray}
\label{masa_inertia_st}
M_{_{\hsn\oplus}} &=& \frac43 \pi \sum_{i=1}^{55} \rho_i (R_{i+1}^3-R_i^3) = M_{\chic{IC}}+M_{\chic{OC}}+M_{M_1}+M_{M_2}+M_C
\nonumber \\
I_{_{\hsn\oplus}}&=& \frac8{15} \pi \sum_{i=1}^{55} \rho_i (R_{i+1}^5-R_i^5) = I_{\chic{IC}}+I_{\chic{OC}}+I_{M_1}+I_{M_2}+I_C
\end{eqnarray} 
where  $M_i$ and $I_i$, $i = IC, OC, M_1, M_2, C$, are the masses and moments of inertia of the major layer 
specified above, which are given by the sums of the contributions of the shells contained in each of these divisions, 
as indicated in Table 1.

To modify the densities of the outer core and the lower and upper mantle we multiply the densities of all 
shells within each of these layers by the respective rescaling factor, $f_{\chic{OC}}, f_{M_1}$, and $f_{M_2}$.
This is done in such a way that neither $M_{_{\hsn\oplus}}$ nor $I_{_{\hsn\oplus}}$ change.
Then,
\begin{eqnarray}
\label{masa_inertia_md}
M_{_{\hsn\oplus}} &=& M_{\chic{IC}}+f_{\chic{OC}} \; M_{\chic{OC}}+ f_{M_1} \; M_{M_1}+ f_{M_2} \; M_{M_2}+M_{\chic{C}}
\nonumber\\
I_{_{\hsn\oplus}}&=&I_{\chic{IC}}+ f_{\chic{OC}} \; I_{\chic{OC}}+f_{M_1} \; I_{M_1}+f_{M_2} \; I_{M_2}+I_{\chic{C}}
\end{eqnarray}
Equating Eqs. \eqref{masa_inertia_st} and \eqref{masa_inertia_md} we obtain the following homogeneous system of 
linear equations:
\begin{eqnarray}
\delta_{\chic{OC}} \; M_{\chic{OC}}+\delta_{M_1} \; M_{M_1}+\delta_{M_2} \; M_{M_2} &=& 0\,,
\nonumber \\
\delta_{\chic{OC}} \; I_{\chic{OC}}+\delta_{M_1} \; I_{M_1}+\delta_{M_2} \; I_{M_2} &=& 0\,,
\end{eqnarray}
where $\delta_{\chic{OC}} = f_{\chic{OC}} -1$, $\delta_{M_1} = f_{M_1} -1$, and $\delta_{M_2} = f_{M_2} -1$ are the 
relative changes of the densities in the outer core, lower mantle, and upper mantle, respectively. Solving 
this system, we can express $\delta_{M_1}$ and $\delta_{M_2}$ as functions of $\delta_{\chic{OC}}$:
\begin{eqnarray}
\label{reldeltaocdeltaM12}
\delta_{M_1}=-\delta_{\chic{OC}} \;\; \frac{\Delta_{M_1}}{\Delta}\,, \qquad \delta_{M_2}=-\delta_{\chic{OC}} \;\; \frac{\Delta_{M_2}}{\Delta}\,,
\end{eqnarray}
where
\begin{eqnarray}
\Delta &=& M_{M_1} \; I_{M_2} - I_{M_1} \; M_{M_2},\nonumber\\
\Delta_{M_1} &=& M_{\chic{OC}} \; I_{M_2} - I_{\chic{OC}} \; M_{M_2},\\
\Delta_{M_2} &=& I_{\chic{OC}} \; M_{M_1} - M_{\chic{OC}}\;I_{M_1}.\nonumber
\end{eqnarray}

\begin{figure}[h!]
\begin{subfigure}[b]{0.5\textwidth}
\includegraphics[width=\textwidth]{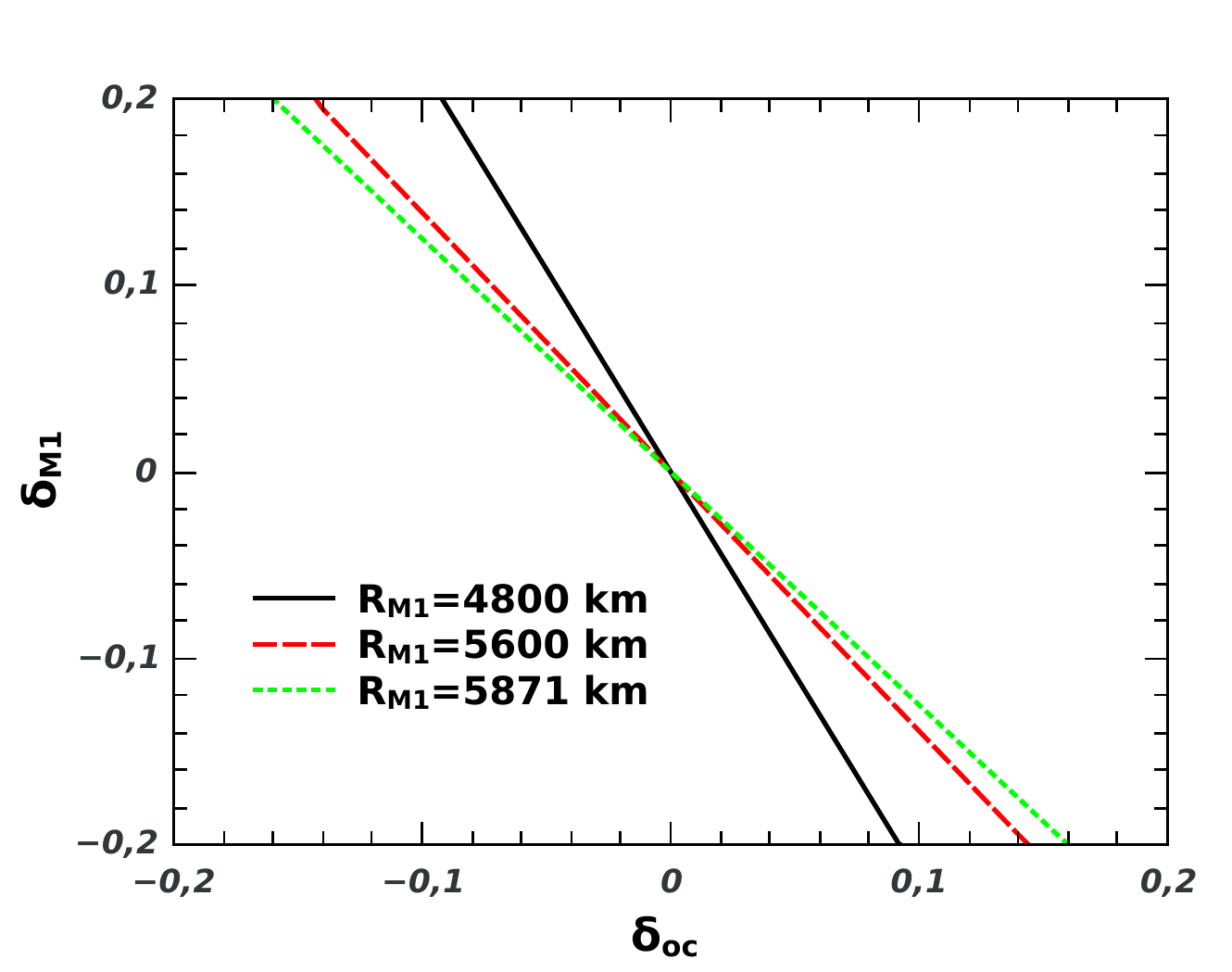}
\caption{}	
%\label{}	
\end{subfigure}
\hfill
\begin{subfigure}[b]{0.5\textwidth}
\includegraphics[width=\textwidth]{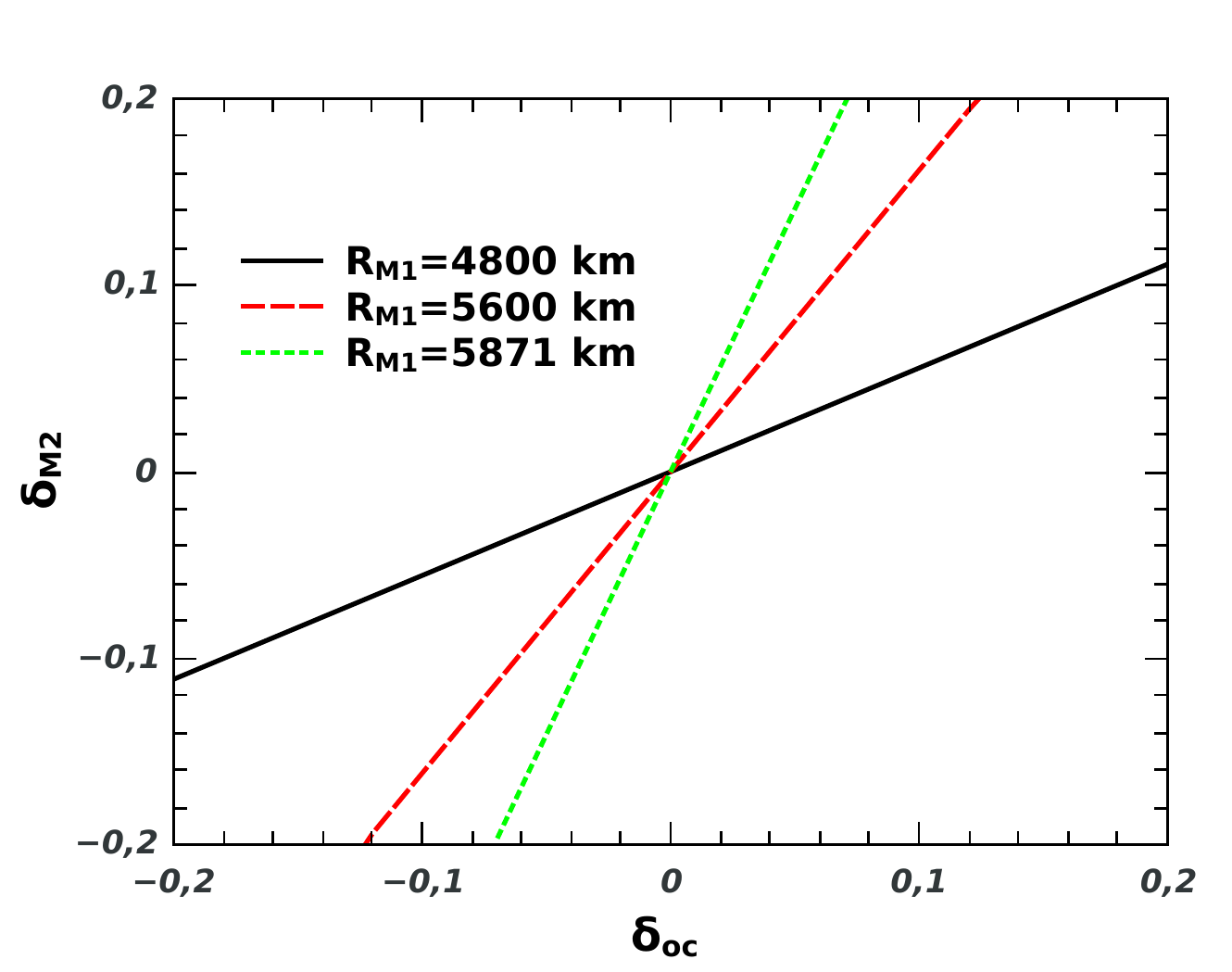}
\caption{}	
%\label{}	
\end{subfigure}
\vskip\baselineskip
\hfill
\caption{Relative changes in the densities of the layers (a) $M_1$ and (b) $M_2$ as a function of the relative change in the density of the outer core, for $R_{M_1}$equal to 
4800 km, 5600 km, and 5871 km.}
\label{fig:dens}
\end{figure}

The value of the radius $R_{M_1}$ set the position of the boundary between the regions $M_1$ and $M_2$ and varying it we can change the number of shells within each of these layers. This, in turn, modifies the 
values of $M_{M_{1,2}}$ and $I_{I_{1,2}}$ and makes the quantities $\delta_{1,2}$  dependent on $R_{M_1}$.
Fig. \ref{fig:dens} shows the relative changes in the densities of layers $M_1$ and $M_2$ as a function of the relative change in the density of the outer core, for three different positions of the boundary between $M_1$ and $M_2$.
\begin{table}
\centering
\begin{tabular}{|c|c|c|}
\hline
\textbf{Parameter} & \textbf{Normal Ordering} & \textbf{Inverted Ordering} \\ \hline
$\Delta m^{2}_{21}$ {[}$\text{eV}^{2}${]} & $7.42\times 10^{-5}$ & $7.42\times 10^{-5}$ \\
$\Delta m^{2}_{31}$ {[}$\text{eV}^{2}${]} & $2.533\times 10^{-3}$ & $-2.437\times 10^{-3}$ \\
$\sin^{2}{\theta_{12}}$ & 0.309 & 0.308 \\
$\sin^{2}{\theta_{13}}$ & 0.0223 & 0.0232 \\
$\sin^{2}{\theta_{23}}$ & 0.561 & 0.564 \\
$\delta/\pi$ & $1.19$ & $1.54$ \\ \hline
\end{tabular}
\caption{Three-neutrino oscillation parameters obtained by averaging the best-fit values of three recent global fits of the current neutrino oscillation data \cite{Capozzi:2020,Esteban:2020cvm,deSalas:2020pgw}.}
\label{tab:constants}
\end{table}
%%%%%%%%%%%%%%%%%%%%%%%%%%%%%%%%%%%%%%%%%%%%%%%%%%%%%%%%%%%%%%%%%
\section{Atmospheric neutrino oscillations} 
\label{sec:osc}

Neutrino oscillations are a well-verified and widely studied phenomenon that has proven beyond any doubt that neutrinos are mixed massive particles. The current experimental and observational data set can be interpreted in terms of the minimal extension of the Standard Model, where the known flavor states $|{\nu}_{\al}\ran (\al = e, \mu, \tau)$ are linear combinations of the states $|{\nu}_i \ran$ with masses $m_i\,(i= 1,2,3)$: 
$
\label{flav&massrel}
|{\nu}_{\al}\ran= \sum_i U^*_{\al i} |{\nu}_i \ran.
$
The coefficients $U_{\al i}$ appear in the leptonic charged current and are elements of a unitary matrix $U$. For for Dirac neutrinos, this matrix can be expressed as
$
\label{eq:Ufactorization}
U= \Oyz\G\Oxz\Oxy\G^*\!,
$
\ni
with 
\bea
\Oxy = 
      \left(
           \ba{ccc}
            c_{12} & s_{12} & 0 \\
           -s_{12} & c_{12} & 0 \\
               0   &    0   & 1 
           \ea
        \right),
&
\ \Oxz = 
      \left(
           \ba{ccc}
            c_{13} &  0  & s_{13} \\
	       0   &  1  &   0    \\
           -s_{13} &  0  & c_{13}
           \ea
        \right),
\nonumber \\
\Oyz = 
      \left(
           \ba{ccc}
            1   &      0   &    0   \\
            0   &   c_{23} & s_{23} \\
            0   &  -s_{23} & c_{23}           \ea
        \right) ,
&
\!\!\! \Gamma = 
      \left(
           \ba{ccc}
            1   &      0   &    0   \\
            0   &      1   &    0   \\
            0   &      0   & {\rm e}^{i\delta}
           \ea
        \right)  ,
\eea

\ni 
where $c_{ij}=\cos\theta_{ij}$ and $s_{ij}=\sin\theta_{ij}$. A value of $\delta$ different form 0 or $\pi$ implies CP-violation
in the leptonic sector of the theory \!\footnote{For Majorana neutrinos there are two additional physical phases, but they are not relevant in neutrino oscillations and are therefore omitted in the analysis of the phenomenon \cite{Bilenky:1980cx}.}\!. 

In addition to the mixing angles $\theta_{ij}$ and the CP-violating phase $\delta$, the oscillations between the three active neutrinos are parametrized by two squared-mass differences: $\D m^2_{21} \equiv m^2_2 - m^2_1$ and  $\D m^2_{32} \equiv m^2_3 - m^2_1$. Five of the parameters ($\theta_{12}, \theta_{13}, \theta_{23}, \D m^2_{21}$, and $|\D m^2_{31}|$), have been determined with remarkable precision ($\thicksim1-5 \%$) by global fits of the data from solar, atmospheric, reactor and long baseline experiments. The issues still pending are: the sign of $\D m^2_{31}$, the $\theta_{23}$ octant, and the determination orfthe phase $\delta$. The sign of $\D m^2_{31}$ characterize the normal ordering (NO), with $m_3 > m_{1,2}$, and the inverted ordering (IO), with $m_3 < m_{1,2}$. The 3$\nu$ oscillation parameters  shown in Table 2 are the mean of the best-fit values for the allowed ranges at $1\sigma$ of the global analyses performed by three groups \cite{Capozzi:2020,Esteban:2020cvm,deSalas:2020pgw}.
 
Neutrinos are produced and detected as flavor eigenstates. Consider a neutrino $\nu_\al$ produced at time $t_0$
that propagates in vacuum. Due to slight mass differences, the phases of the mass eigenstate components of the original flavor state change at different rates and, due to this, the flavor content of the neutrino beam oscillates along the trajectory. When neutrinos propagate in a medium, the coherent forward scattering of neutrinos with electrons is different for  $\nu_e$ and $\nu_{\mu,\tau}$, resulting in different refraction indexes for the electron neutrino and the other flavors. As a consequence, neutrino oscillations can be significantly modified  in matter compared to oscillations in vacuum and new resonance enhancement effects appear. These effects are sensitive to the density and composition of the medium and we will take advantage of this in order to examine the inner parts of our planet by means of the oscillations of atmospheric neutrinos in the Earth.
Atmospheric neutrinos have played a very important role in the study and characterization of the phenomena of neutrino oscillations. They are generated around the Earth as decay products in hadronic showers that result from collisions of cosmic rays with nuclei in the upper atmosphere. This provide a continuous source of neutrinos spanning a very wide range of energies and travelled distances before detection. On this work, we concentrate on those with energies in the range of 1-10 GeV and different nadir angles.

Let a neutrino $\nu_\al$ that enters the solid terrestrial matter at time $t_0$. At any time $t > t_0$ the state of the system $|\psi(t)\ran$ can be expressed as $|\psi(t)\ran = \hat {\mc U}(t, t_0)|\psi(t_0)\ran$, where $|\psi(t_0)\ran = |\nu_{\al}\ran$ and $\hat{\mc U}(t, t_0)$ is the evolution operator. The probability of having a neutrino of flavor $\beta$ inside the Earth, at a distance $\ell \simeq t - t_0 \hs{0.02 cm} (\hslash = c = 1)$ from the entry point, is
\be
\label{oscprob}
P_{{\nu_\al} \to {\nu_\b}}(\ell) = |\,{\mc U}_{\b \al}(\ell)|^2\,,
\ee
where the probability amplitude 
$
\label{matrixelem}
{\mc U}_{\b \al}(\ell)= \lan \nu_{\b}|\,\hat {\mc U}(\ell)|\nu_{\al}\ran
$
is an element of the $3 \times 3$ unitary matrix $\mc{U}(\ell)$ representing the evolution operator in the flavor basis.
\begin{figure}[h!]
\centering
\includegraphics[width=0.45\textwidth]{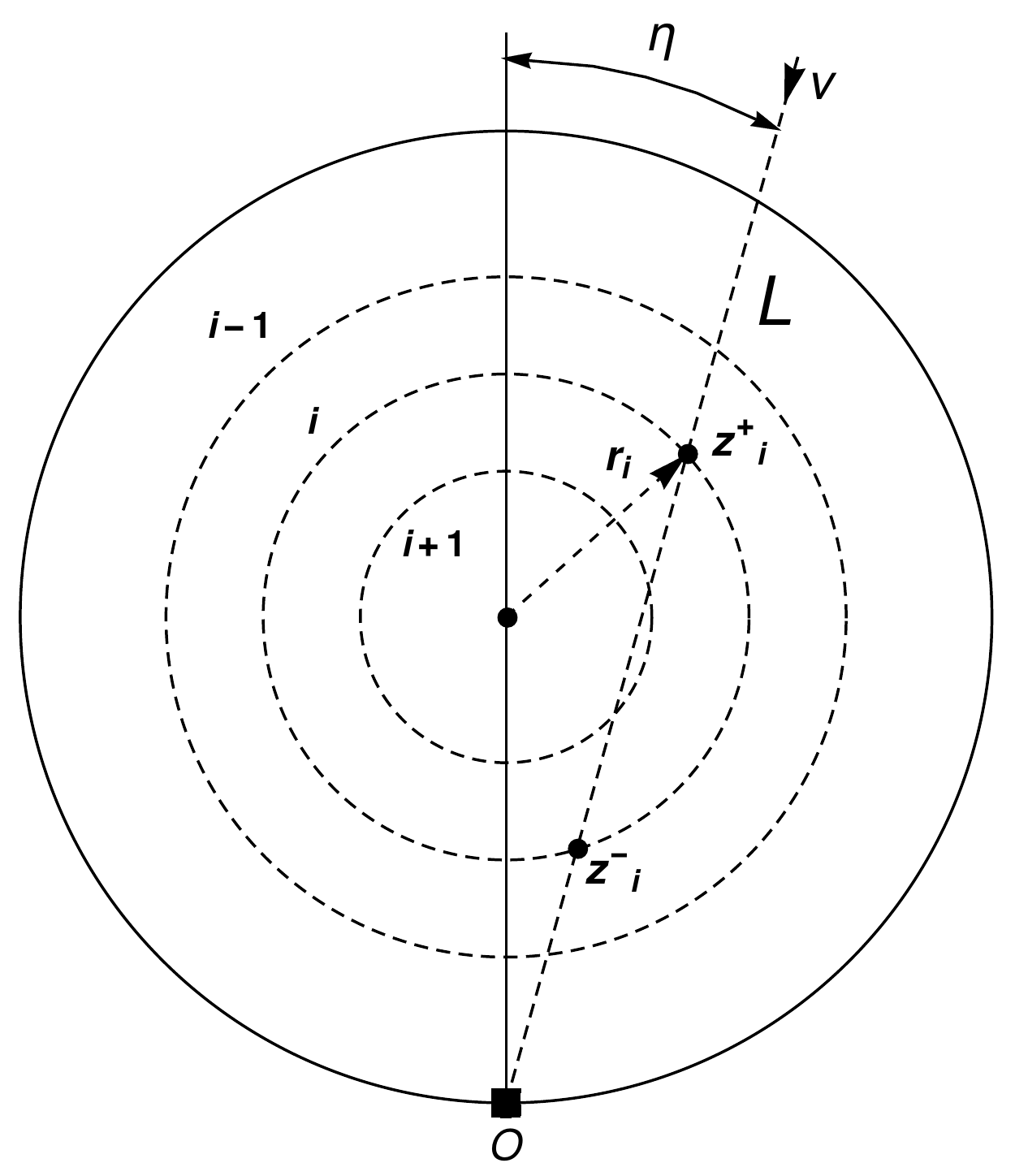}	
%\label{}	
\hfill
\caption{Neutrino path through the Earth.}
\label{fig:earth}
\end{figure}
Rather than solving for $\mc{U}(\ell)$ directly, it proves to be more convenient to determine the evolution operator in the basis of the mass eigenstates and then transform it to the flavor basis using the relation
\be
\mc{U}(\ell) = \tilde U \tilde{\mc{U}}(\ell) \tilde{U}^{\dagger} ,
\ee
where $\tilde U =  \Oyz\G\Oxz\Oxy$.

The matrix $\tilde{\mc{U}}(\ell)$ obeys the equation
\!\footnote{
The operator $\tilde{\mc{U}}(\ell)$ evolves the wave function $\tilde{\Phi}(\ell) = {\G^*}\Phi(\ell)$, with $\Phi^{\mtt T}(\ell)  = 
\left(\phi_1(\ell), \phi_2,(\ell), \phi_3(\ell)\right)$, where
 $\phi_i(\ell), i =1,2,3$ are the amplitudes of the mass eigenstates.}
\be
\label{Scheq}
i \frac{d}{d \ell}\tilde{\mc{U}}(\ell) = \tilde H (\ell) \tilde{\mc{U}}(\ell), \qquad \tilde{\mc{U}}(0) = I,
\ee
with 
\be  \
\label{eq:Hmatter}
\tilde H (\ell) =  H_{\rm 0}  + {\mtt V}(\ell) {\mc O}^{\mtt T} Y {\mc O}\,.
\ee
Here, $H_0 = {\rm diag}\{0, \D_{21}, \D_{31}\}$  and $Y = {\rm diag}(1,0,0)$ are diagonal matrices
and ${\mc O}^{\mtt T}$ is the transpose of the orthogonal matrix ${\mc O} = \Oxz\Oxy$. 
Note that the real and symmetric matrix ${\mc O}^{\mtt T} Y {\mc O} $ does not depend on $\th_{23}$ 
nor $\de$. Explicitly,
\be 
\label{eq:matrixW}
 {\mc O}^{\mtt T} Y {\mc O}= \left(\!
\begin{array}{ccc}
c_{\xz}^2c_{\xy}^2 & c_{\xy}s_{\xy}c_{\xz}^2 & c_{\xy}c_{\xz}s_{\xz} \\
c_{\xy}s_{\xy}c_{\xz}^2 & s_{\xy}^2c_{\xz}^2 & s_{\xy}c_{\xz}s_{\xz} \\
c_{\xy}s_{\xz}c_{\xz} & s_{\xy}c_{\xz}s_{\xz} & s_{\xz}^2
\end{array}
\!\right)\!.
\ee
%
%
%\begin{table}
%\centering
%\begin{tabular}{ccc}
%\hline
%\textbf{Parameter} & \textbf{Normal Ordering} & \textbf{Inverted Ordering} \\ \hline
%$\Delta m^{2}_{21}$ {[}$\text{eV}^{2}${]} & $7.42\times 10^{-5}$ & $7.42\times 10^{-5}$ \\
%$\Delta m^{2}_{31}$ {[}$\text{eV}^{2}${]} & $2.533\times 10^{-3}$ & $-2.437\times 10^{-3}$ \\
%$\sin^{2}{\theta_{12}}$ & 0.309 & 0.308 \\
%$\sin^{2}{\theta_{13}}$ & 0.0223 & 0.0232 \\
%    $\sin^{2}{\theta_{23}}$ & 0.561 & 0.564 \\
%$\delta/\pi$ & $1.19$ & $1.54$ \\ \hline
%\end{tabular}
%\caption{Parameters of the three-neutrino oscillations obtained by averaging the best-fit values of recent global fits of the current neutrino oscillation data \cite{Capozzi:%2020,Esteban:2020cvm,deSalas:2020pgw}.}
%\label{tab:constants}
%\end{table}
The first term in  Eq. \eqref{eq:Hmatter} is the Hamiltonian that drives the flavor evolution in vacuum, while
the second term accounts for the matter effects. For antineutrinos, the sign of the second term is reversed
and the matrix $\G$ is replaced by its complex conjugate. 

In normal matter (n, p, e)
\be
{\mtt V}(\ell) =  \sqrt 2 G_F n_e(\ell)\,,
\ee
where $n_e(\ell)$ is the electron number density and $G_F$ is the Fermi constant. 
The electron number density depends on both the matter density $\rho(\ell)$ and the chemical and isotopic composition 
of the medium:		
\be
\label{elecdensity}
n_e (\ell) =\dfrac{{\rho}(\ell)}{m_{\rm u}} \dfrac{Z}{A}( \ell)\,,
\ee
where $m_{\rm u} = 931.494$ MeV is the atomic mass unit and
$
Z/A =  \sum_{\lam} \digamma_{\!\!\lam} (Z/A)_{\lam} .
$
The summation runs over all the chemical elements present in the medium and $(Z/A)_{\lam}$ denotes the ratio between the atomic number $Z_{\lam}$ and the atomic mass $A_{\lam}$ of the element that contributes a fraction $\digamma_{\!\!\lam} $ to the mass at a given position.

The relevant quantities for us are the oscillation probabilities for ${\nu_\mu} (\bar{\nu}_\mu) \to {\nu_{\mu.\tau}} (\bar{\nu}_{\mu,\tau})$
and ${\nu_e} (\bar{\nu}_e) \to {\nu_{\mu,\tau}} (\bar{\nu}_{\mu,\tau})$, which are required to compute the $\mu$-like events produced in a detector by 
``upward'' atmospheric neutrinos after traveling a distance $L = 2 R_{_{\hsn\oplus}} \cos \eta $ through the Earth (see Eq. \eqref{eq:numeroeventos}).
According to the previous considerations, because of the dependence of these probabilities on the potential ${\mtt V}(\ell)$, atmospheric neutrinos are sensitive to changes in the density and composition of the traversed layers. In what follows, we examine the feasibility of applying such an effect to obtain  meaningful information about the deepest part of the Earth.  With this in mind, we calculate $\tilde{\mc{U}}(L)$ for the Earth modeled as a sphere made up of 55 concentric spherical shells with different (constant) densities, as described in Sec. \eqref{sec:structure}. 

The complete evolution operator can be expressed as the product of the evolution operators for the consecutive shells through which the neutrinos pass on their way to the detector:
\begin{eqnarray}
\label{fullevolop}
\tilde{\cal U}(L) = \prod_{j=1}^{2 j_m-1}\tilde{\cal U}_j(L_j)\,,
\end{eqnarray} 
where $L_J$ are the distances in each shell, such that
$L = \sum_j L_j$
and 
\be
\tilde{\cal U}_j(L_j) = 
\exp(-i{\tilde H}_j L_j)\,.
\ee
The effective potential in ${\tilde H}_{j}$ takes the fixed value $ {\mtt V}_{j} = \sqrt{2}\hsp G_{\hs{-.03cm}F} n^{j}_e,$ where  $n^{j}_e$ denotes the constant number density of electrons in shell $j$.  Since the Hamiltonians for the different layers do not generally commute between them, the exponential factors  in Eq. \eqref{fullevolop} must be in the prescribed order. Each of these factors can be evaluated by applying the Cayley-Hamilton theorem which allows us to convert the infinity series into a polynomial:
$
\tilde{\mc U}_{j}(L_{j})  = a^j_0 I+a^j_1 {\tilde H}_{\ell} + a^j_2 {\tilde H}_{\ell}^2 \,,
$
where the coefficients are functions of the eigenvalues of ${\tilde H}_{j}$ (which coincide with those of $H_{j}$) 
\cite{Ohlsson:1999xb, DOlivo:2020ssf}.

\begin{table}[h]
\centering
\begin{tabular}{@{}|c|c|c|c|@{}}
\toprule
\textbf{Set} & \textbf{$f_{oc}$} & \textbf{$f_{(Z/A)_{oc}}$} & \textbf{$f_{(Z/A)_{M_1}}$} \\ \midrule
I   & 1    & 1    & 1    \\
II  & 1    & 1.01 & 1    \\
III & 1.01 & 1    & 1    \\
IV  & 1    & 1    & 1.01 \\ \bottomrule
\end{tabular}
\caption{Different sets of density and composition.}
\label{tab:sets}
\end{table}

The distances $L_j$ are functions of the nadir angle $\eta$. According to Fig. \ref{fig:earth}, they 
can be determined as 
$ L_j=\vert \tilde{z}_{j+1} - \tilde{z}_j \vert$,
where
\begin{equation}
 \tilde{z}_j = \begin{cases}
z^+_j\,,  & 1 \le j \le j_m\,,\\
z^-_{2j_m +1 -j}\,, & j_m + 1 \le j \le 2j_m\,,
\end{cases}
\end{equation}
\begin{equation}
z^{\pm}_ j = R_{_{\hsn\oplus}} \cos\eta\pm \sqrt{r_j^2-(R_{_{\hsn\oplus}} \sin\eta)^2}\,.
\end{equation}
\begin{figure}[h!]
\centering
\begin{subfigure}[b]{0.5\textwidth}
\includegraphics[width=\textwidth]{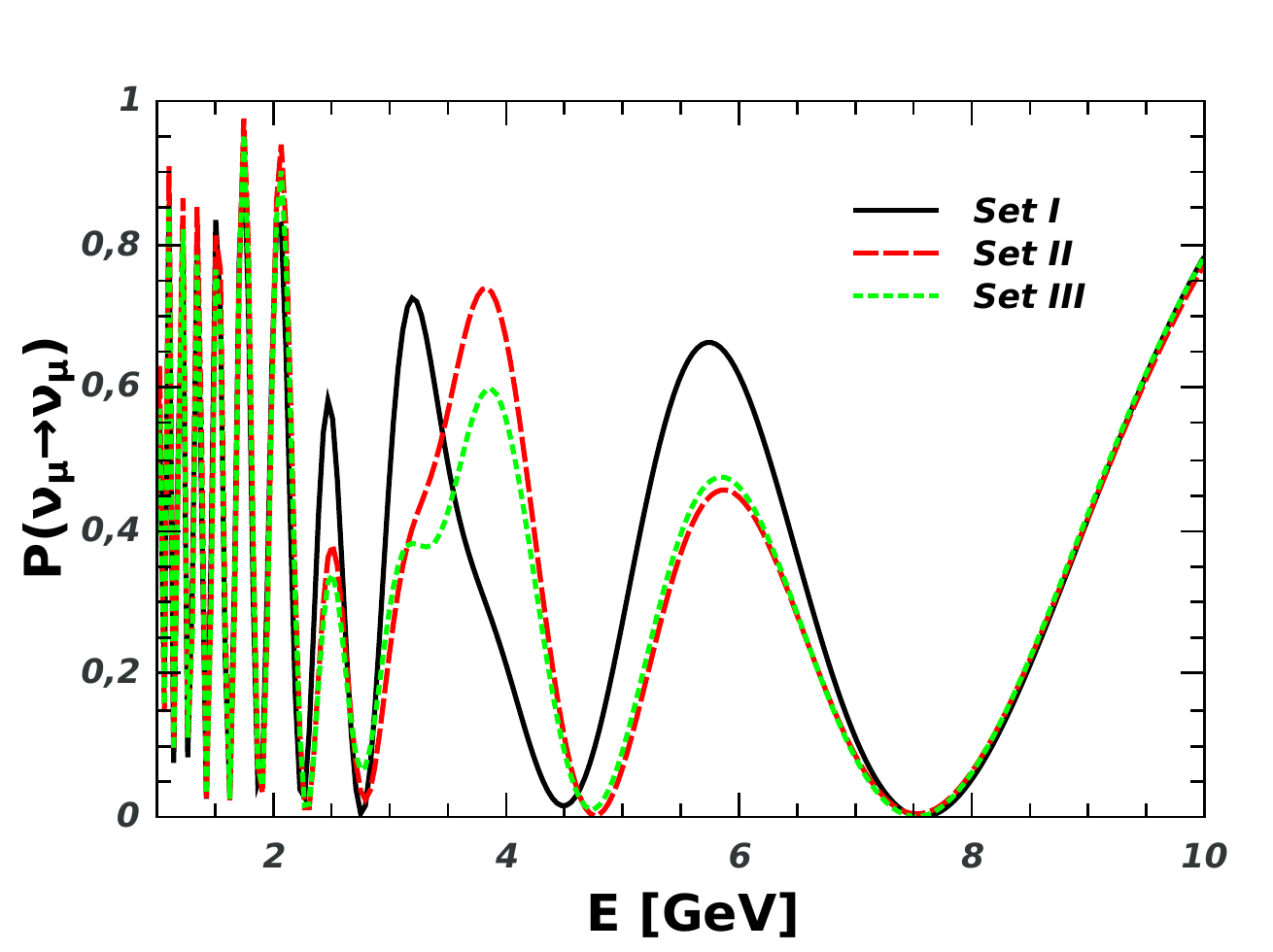}
\caption{NO and $\eta = 20^{\circ}$}	
%\label{}	
\end{subfigure}\begin{subfigure}[b]{0.5\textwidth}
\includegraphics[width=\textwidth]{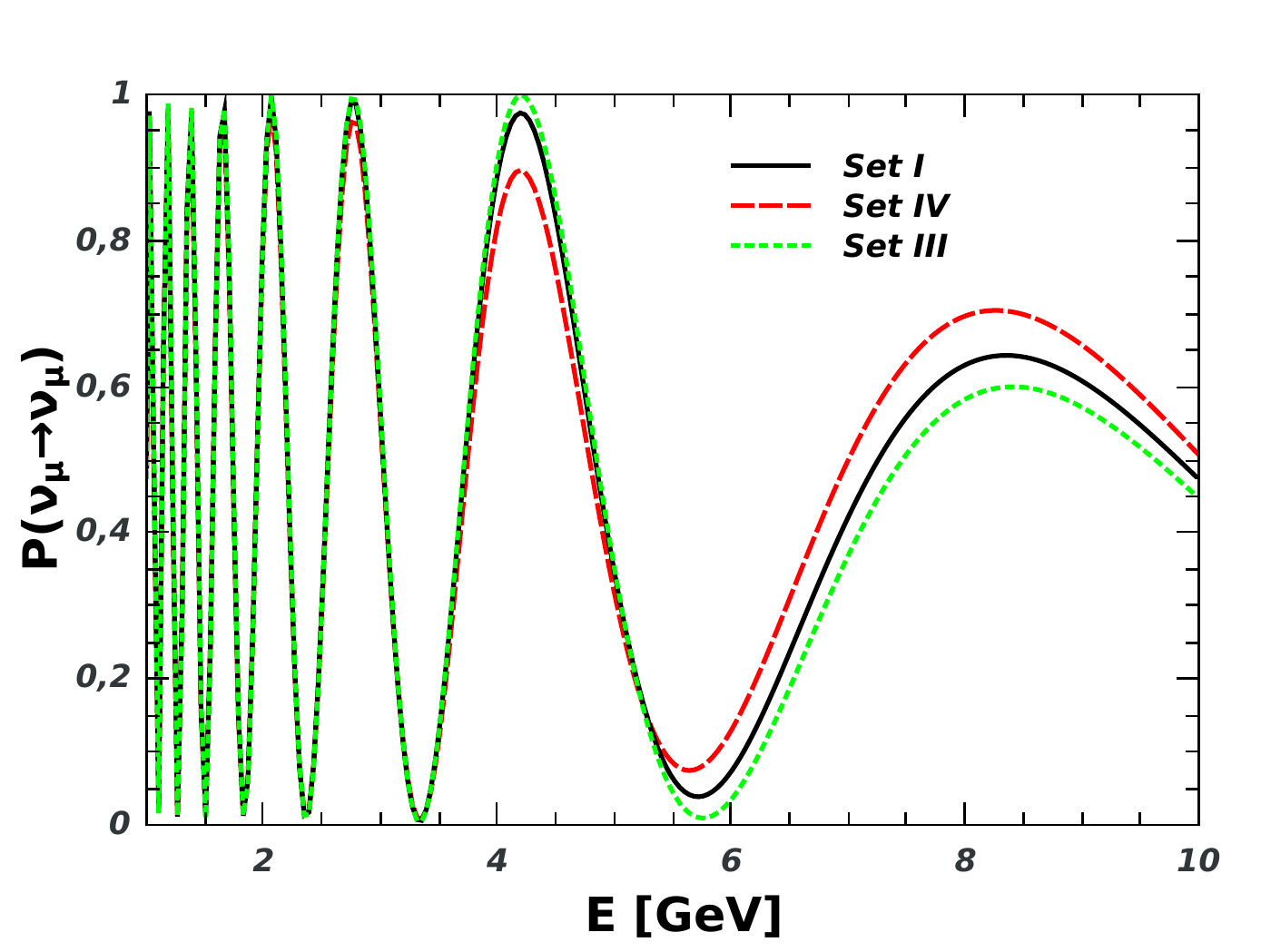}
\caption{NO and $\eta = 50^{\circ}$}	
%\label{}	
\end{subfigure}
\caption{Survival probability $P_{{\nu_\mu} \to {\nu_\mu}}$ as a function of the neutrino energy, for normal ordering (NO) and nadir angle $\eta$ equal to (a) $20^{\circ}$
and (b) $50^{\circ}$\!.}
\label{fig:mu--mu}
\end{figure}
\begin{figure}[h!]
\centering
\begin{subfigure}[b]{0.5\textwidth}
\includegraphics[width=\textwidth]{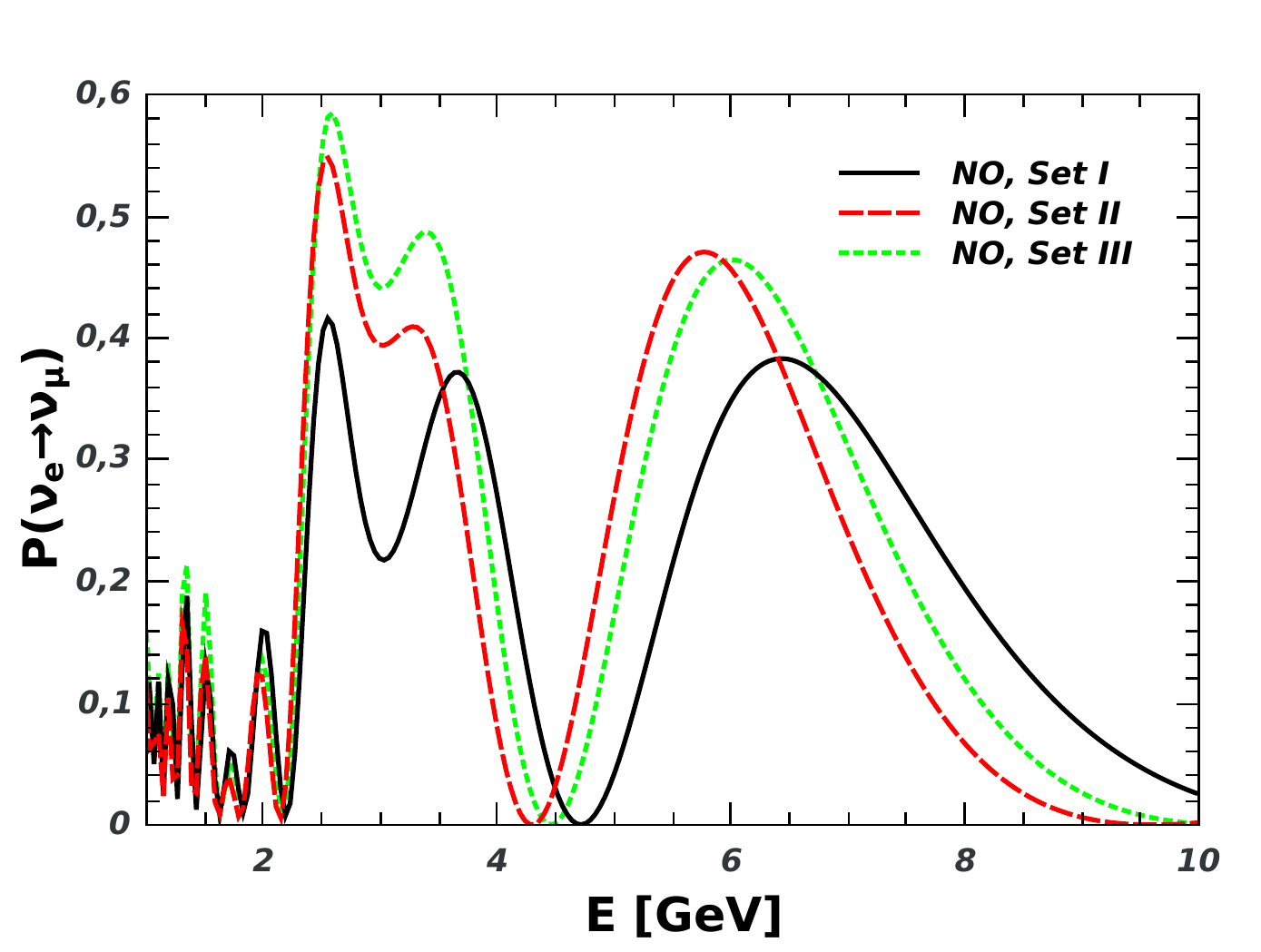}
\caption{NO and $\eta = 20^{\circ}$}	
%\label{}	
\end{subfigure}\begin{subfigure}[b]{0.5\textwidth}
\includegraphics[width=\textwidth]{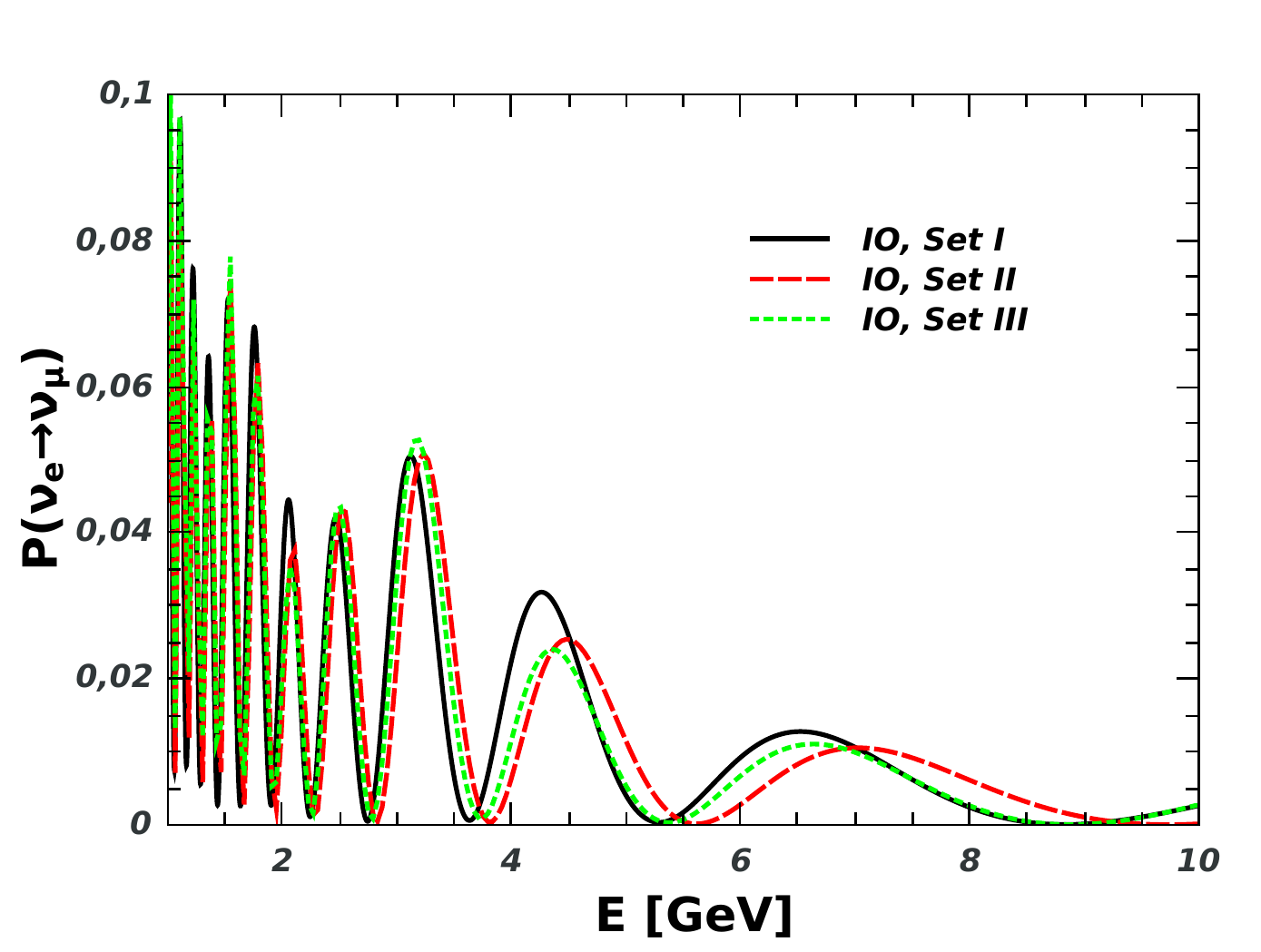}
\caption{IO and $\eta = 20^{\circ}$}	
%\label{}	
\end{subfigure}
%\hfill
\begin{subfigure}[b]{0.5\textwidth}
\includegraphics[width=\textwidth]{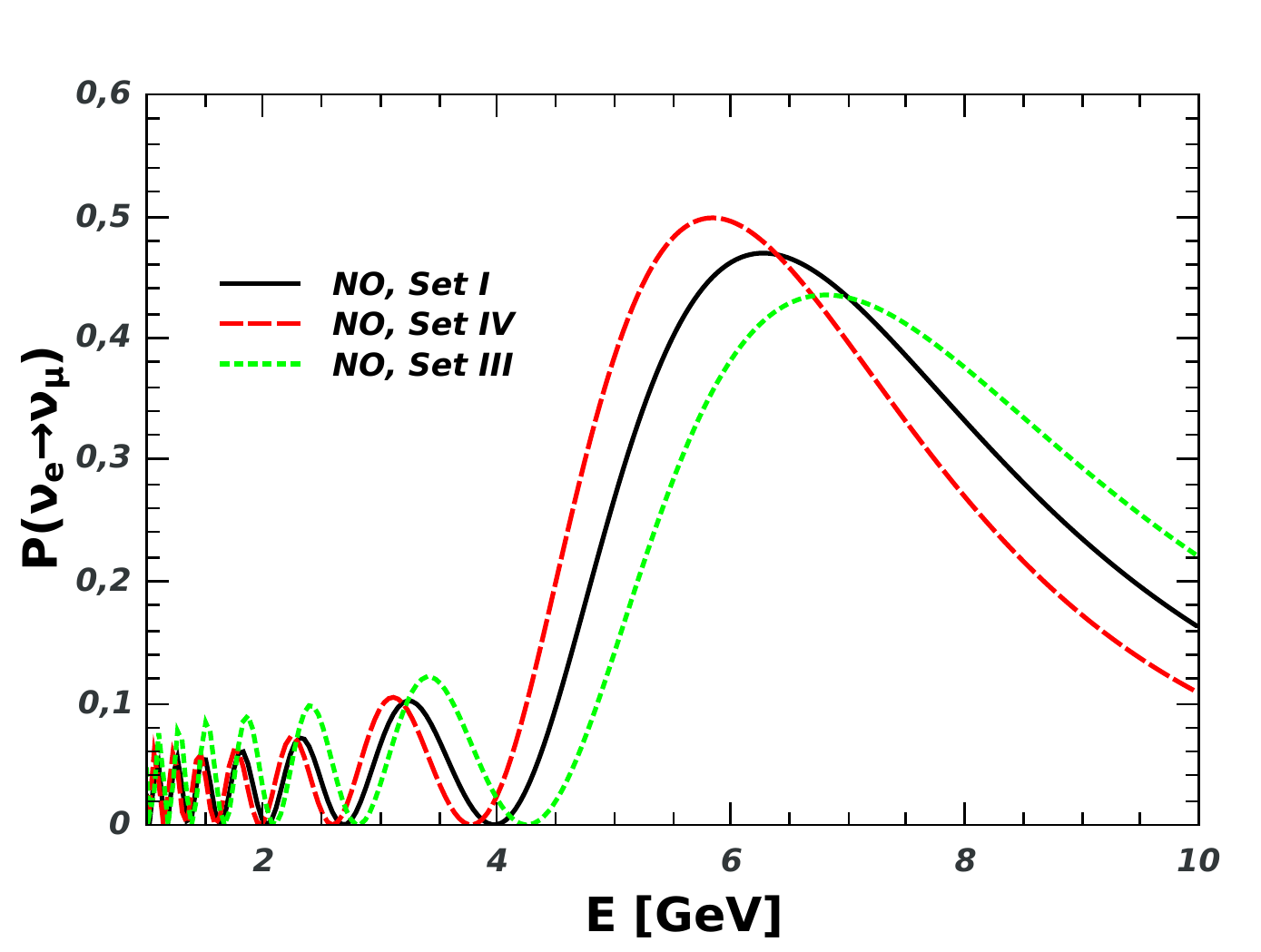}
\caption{NO and $\eta = 50^{\circ}$}	
%\label{}	
\end{subfigure}\begin{subfigure}[b]{0.5\textwidth}
\includegraphics[width=\textwidth]{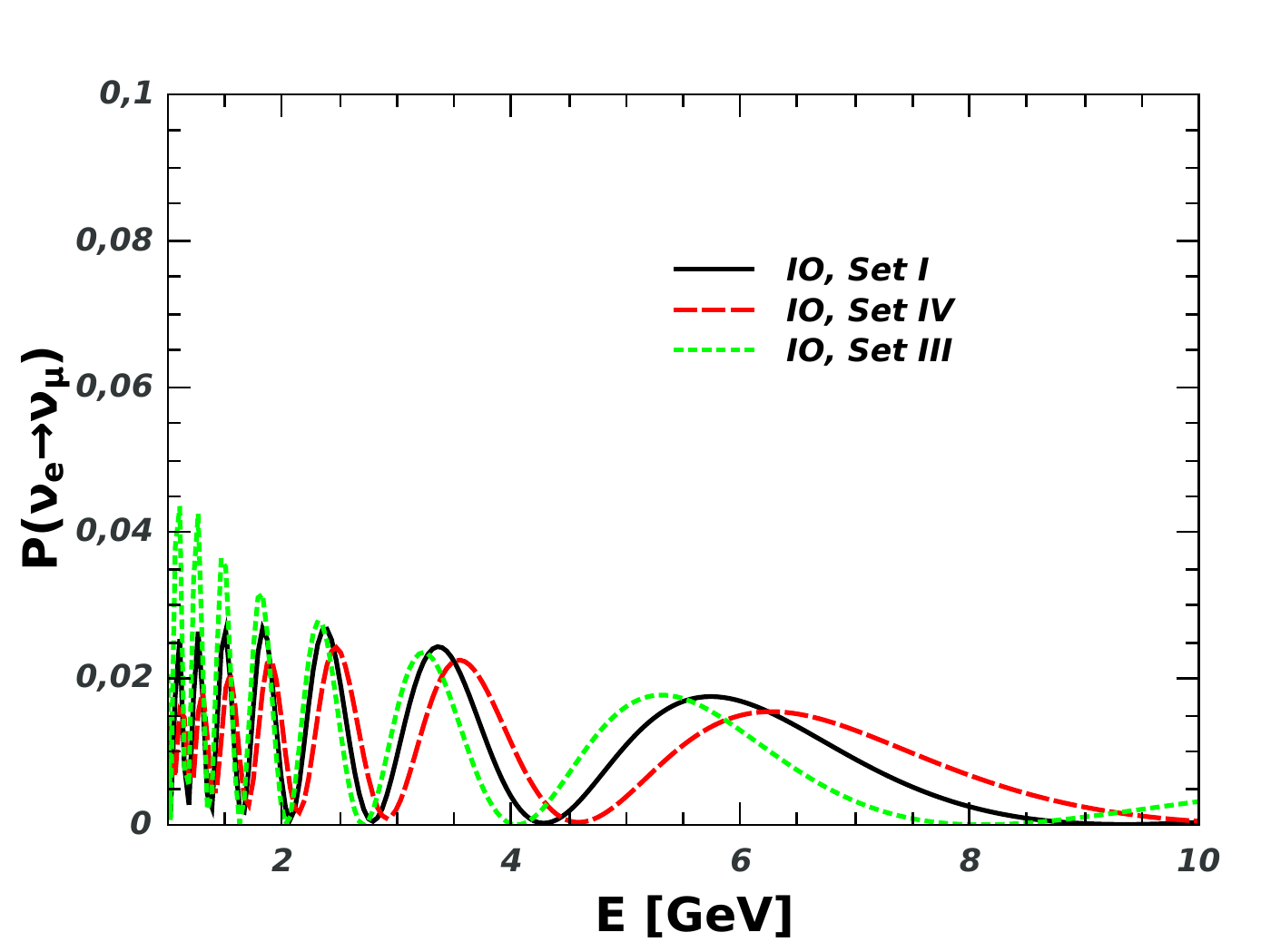}
\caption{IO and $\eta = 50^{\circ}$}	
%\label{}	
\end{subfigure}
\caption{Transition probability $P_{{\nu_e} \to {\nu_\mu}}$ as a function of the neutrino energy, for normal ordering (NO) and inverted ordering (IO) and nadir angle $\eta$ equal to (a,b) $20^{\circ}$ and (c,d) $50^{\circ}$\!.}
\label{fig:mu--e}
\end{figure}
To exemplify the effect that changes in composition and density have on the probabilities of flavor oscillations, in Figs.~\ref{fig:mu--mu} and \ref{fig:mu--e} we show the probabilities for $\nu_{\mu} \to \nu_{\mu}$ and $\nu_{e}\to \nu_{\mu}$ as functions of the neutrino energy, for two different angles of incidence, both for NO and IO.  In the next section, we use these probabilities in the calculation of  the $\mu$-like events produced by atmospheric neutrinos arriving at the detector after passing through the internal regions of the Earth.  

%%%%%%%%%%%%%%%%%%%%%%%%%%%%%%%%%%%%%%%%%%%%%%%%%%%%%%%%%%%%%%%%%%%%%%%%%%%%%%%%%%%%%%%%%%%%%%%%%%%%%%%%%%%
\section{Neutrino events and test of Earth's composition}
\label{sec:simu}

Let us consider a generic detector with a mass of water or ice containing a number $n_{\chic N}$ of nucleons as a target. Such a detector, like the planned ORCA, will efficiently detect the Cherenkov radiation emitted along the path of $\mu^{\pm}$ produced by the interactions of $\nu _ {\mu}$ and $\bar\nu_{\mu} $ with the nucleons in the instrumented volume or near it. In addition, it will be posible to detect showers due to the $e$ or $\tau$ produced by the associated neutrinos \cite{Stavropoulos:2021hir,Kalaczynski:2021ytv,Bourret:2017tkw}. 

To generate a data set that simulates the actual observations, with the inevitable statistical fluctuations, we perform a numerical calculation of the events due to neutrinos.  Our observable is the number of $\mu$-like events ${\mathcal N}_{\mu}$ and in what follows we use it to test different hypothesis about the composition and density of the outer core and the lower part of the mantle. 
To identify the most sensitive regions, we first do a scan dividing the cone under the detector into a series of angular and energy bins and calculate the number of $\mu$-like events within given angular and energy intervals:
\begin{eqnarray} 
\label{eq:numeroeventos}
 {\mathcal N}_{\mu}= n_{\chic N} T \int_{E_{\text{min}}}^{E_{\text{max}}} \! \dd{E} \int_{\eta_{\text{min}}}^{\eta_{\text{max}}} \dd{\Omega}
\left[\sigma^{cc}_{\nu_{\mu}}(E)\left(P_{\nu_{\mu}\rightarrow \nu_{\mu}} {\dfrac{d\Phi}{dE}}^{\!\!\nu_{\mu}}
\!+ P_{\nu_{e}\rightarrow \nu_{\mu}}{\dfrac{d\Phi}{dE}}^{\!\!\nu_{e}}\right) +\right.
\nonumber \\
\left. \sigma^{cc}_{\bar \nu_{\mu}}(E)\left(P_{\bar \nu_{\mu}\rightarrow \bar \nu_{\mu}} {\dfrac{d\Phi}{dE}}^{\!\!\bar \nu_{\mu}}
\!+ P_{\bar \nu_{e}\rightarrow \bar \nu_{\mu}} {\dfrac{d\Phi}{dE}}^{\!\!\bar \nu_{e}} \right) + \right.
\nonumber \\
\left. \sigma^{cc}_{\nu_{\tau}}(E) Br_{\tau \rightarrow \mu} \left(P_{\nu_{\mu}\rightarrow \nu_{\tau}} {\dfrac{d\Phi}{dE}}^{\!\! \nu_{\mu}}
\!+ P_{\nu_{e}\rightarrow \nu_{\tau}} {\dfrac{d\Phi}{dE}}^{\!\! \nu_{e}} \right) + \right.
\nonumber \\
\left. \sigma^{cc}_{\bar \nu_{\tau}}(E) Br_{\bar\tau \rightarrow \bar\mu} \left(P_{\bar \nu_{\mu}\rightarrow \bar \nu_{\tau}} {\dfrac{d\Phi}{dE}}^{\!\!\bar \nu_{\mu}}
\!+ P_{\bar \nu_{e}\rightarrow \bar \nu_{\tau}} {\dfrac{d\Phi}{dE}}^{\!\!\bar \nu_{e}} \right)
 \right],
\end{eqnarray}
where $T$ is the detection time. The fluxes of atmospheric neutrinos and antineutrinos were taken from Ref. \cite{Atmnu:1996}, 
while the charged-current cross sections for the ${\nu}_{\mu}(\bar{\nu}_{\mu})$- nucleon and ${\nu}_{\tau}(\bar{\nu}_{\tau})$-nucleon scatterings, in the considered range of neutrino energies ($ 1-10$ GeV), are given approximately by \cite{Formaggio:2013kya}:
\begin{equation}
\begin{aligned}
\sigma^{cc}_{\nu_{\mu}}(E) &\simeq 0.75 \times 10^{-38} \left({E}/{\rm GeV}\right) {\rm cm}^2\,, \\
\sigma^{cc}_{\bar \nu_{\mu}}(E) &\simeq 0.35 \times 10^{-38} \left({E}/{\rm GeV}\right) {\rm cm}^2\,,\\
\sigma^{cc}_{\nu_{\tau}}(E) &\simeq 0.13 \times 10^{-38} \left({E}/{\rm GeV}\right) {\rm cm}^2\,, \\
\sigma^{cc}_{\bar \nu_{\tau}}(E) &\simeq 0.05 \times 10^{-38} \left({E}/{\rm GeV}\right) {\rm cm}^2 \,.
\end{aligned} 
\end{equation} 
In Eq. \eqref{eq:numeroeventos}, $Br_{\tau \rightarrow \mu} = Br_{\bar\tau \rightarrow \bar\mu} \simeq 0.17$ are the branching ratios of the decays $\tau \rightarrow \mu {\bar \nu}_\mu \nu_\tau$ and $\bar\tau \rightarrow {\bar\mu}{\nu_\mu} {\bar\nu}_\tau$. The dependence of ${\mathcal N}_{\mu}$ on the density and composition of the medium is incorporated through the oscillation probabilities, which are calculated from the expressions given in Sec. \ref{sec:osc}. It is understood that these probabilities are evaluated at $L$.
The values of the oscillation parameters are those given in Table \ref{tab:constants}. To estimate the impact of current uncertainties on the oscillation parameters, in addition to the best fits, we consider their values at the extremes of the 1$\sigma$ ranges \cite{Capozzi:2020,Esteban:2020cvm,deSalas:2020pgw}. Thus, the muon numbers were also calculated by evaluating the oscillation probabilities at those values of a given parameter and keeping the best fits for the rest. We find that errors in the oscillation parameters introduce only few percent variations in the number of muon events and have little effect on our analysis.

To determine the angular and energy intervals where ${\mathcal N}_{\mu}$ is more sensitive to changes in the compositions of either the outer core or the lower mantle we introduce the quantity 
\begin{equation}
\label{eq:upsilon}
{\Upsilon}(E,\eta) \equiv \abs{1 - \dfrac{\mathcal{N}_{\mu}^{(Z/A)}}{\mathcal{N}_{\mu}^0}} \times 100\,,
\end{equation}
which gives the percentage difference between the number of events for the standard composition $\mathcal{N}_{\mu}^{0}$ and the number of events for a different composition $\mathcal{N}_{\mu}^{\rm (Z/A)}$. The radii of the layers and the densities of the shells in the layers are those given from the PREM.

\begin{figure}[h!]
\centering
\begin{subfigure}[b]{.50\textwidth}
\includegraphics[width=\textwidth]{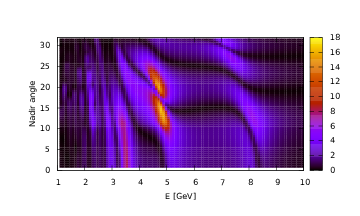}	
\caption{Normal Ordering}
\label{fig:SNO_101_1}
\end{subfigure}\begin{subfigure}[b]{.50\textwidth}
\includegraphics[width=\textwidth]{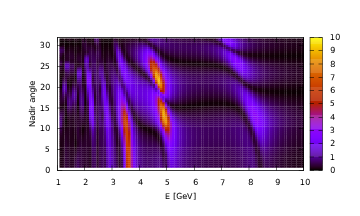}
\caption{Inverse Ordering}
\label{fig:SIO_101_1}
\end{subfigure}
\caption{Level surfaces of the function $\Upsilon(E,\eta)$ (Eq. \eqref{eq:upsilon}) in the ($E,\eta$) plane, for a $1\%$ change in the composition of the outer core.}
\label{fig:sens_OC}
\end{figure}

\begin{figure}[h!]
\centering
\begin{subfigure}[b]{.50\textwidth}
\includegraphics[width=\textwidth]{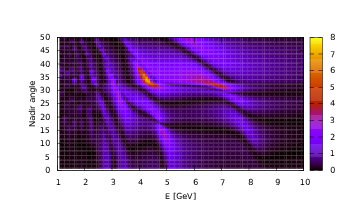}	
\caption{Normal Ordering}
\label{fig:SNO_1_101}
\end{subfigure}\begin{subfigure}[b]{.50\textwidth}
\includegraphics[width=\textwidth]{SNO_1_101.png}	
\caption{Inverse Ordering}	
\label{fig:SIO_1_101}	
\end{subfigure}
\caption{Level surfaces of the function $\Upsilon(E,\eta)$ (Eq. \eqref{eq:upsilon}) in the ($E,\eta$) plane, for a $1\%$ change in the composition of layer $M_1$.}
\label{fig:sens_D2}
\end{figure}

Figs.~\ref{fig:sens_OC} and \ref{fig:sens_D2} show the level surfaces of $\Upsilon$ in the ($E,\eta$) plane 
for a change of $1\%$ in the composition of the outer-core and the composition of the $M_1$ region, respectively, for both NO and IO.  From the figures, it is apparent that in the case of the outer core the most sensitive regions correspond to energies around 
5 GeV and angles compatible with the shadow of the outer core and $M_1$.  On the other hand, for the lower mantle the angular region has to be increased and the energy shifts to slightly lower values. The deviations in the number of events are considerably more pronounced for NO, mainly because for IO the resonance effects in matter occur for antineutrinos, whose charged cross-section is about two times smaller than that of neutrinos. 

To simulate the number of events observed by a km$^3$ detector as ORCA, we follow the same procedure implemented in \cite{DOlivo:2020ssf}, which we reproduce here for completeness.
In order to perform the Monte Carlo simulation, based on the previous results, we consider events with $4 \,\text{GeV} < E < 11\, \text{GeV}$ and $10\degree<\,\eta\,<60\degree$. Both of these intervals are divided into $200$ subintervals. Every pseudo experiment is made up by tossing, in each square bin of the grid, a number of Poisson distributed events with the mean value equal to ${\mathcal N}_{\mu}$ as given by Eq.~(\ref{eq:numeroeventos}). Thus, each of the $n_{exp}$ pseudo experiments consists of 200 $\times$ 200 numbers corresponding to events, one for each bin. This sample of events are then distributed in angle and energy. We call them the  {\it true events} and suppose that they are distributed according to the probability distribution function (pdf)  $f^{\,i_{exp}}_t(E,\eta)$, ${i_{exp}} = 1, \cdots, n_{exp}$. For this function we take the normalized histograms constructed by means of the Monte Carlo simulation. The fractional number of true events in bin $(i,j)$ for the $i_{exp}$-th experiment is given by the integral of $f^{\,i_{exp}}_t(E,\eta)$ over the energy and angle intervals of the bin $(i,j)$.

To obtain a realistic distribution of events we must allow for a limited resolution of the detector, both in energy and angle. The net effect is the redistribution of the true events ("smearing") in the grid bins, which is implemented by folding the true distribution with a resolution function $\mathcal{S}(E_{\mtt o}, \eta_{\mtt o} \vert E, \eta)$. We assume a Gaussian smearing and write
\begin{eqnarray} 
\label{smearing}
\mathcal{S}(E_{\mtt o}, \eta_{\mtt o} \vert E, \eta)=\dfrac1{2\pi \Delta \eta(E) \Delta E(E)}\exp\left(-\dfrac{(\eta - \eta_{\mtt o})^2}{2\Delta\eta(E)^2}\right) \exp\left(-\dfrac{(E-E_{\mtt o})^2}{2\Delta E(E)^2}\right),
\end{eqnarray}  
with the detector characterized by the angular and energy resolutions $\Delta \eta(E) = \alpha_{\eta}/\sqrt{E/\text{GeV}}$ and $\Delta E(E) = \alpha_{E} \, E$, respectively. For the values of the parameters $\alpha_{\eta}$ and $\alpha_{E}$ we consider different situations according to the discussion in  Ref. \cite{Rott:2015kwa}. To keep our analysis as simple as possible we assume a detection efficiency of $100\%$. 
When convoluted with the true events the kernel in Eq. \ref{smearing} gives us what we call the {\it observed events}. That is, $\mathcal{S}(E_{\mtt o}, \eta_{\mtt o} \vert E, \eta)$ represents the conditional pdf for the measured values to be $(E_{\mtt o}, \eta_{\mtt o})$ if the true values were $(E, \eta)$. Since the event is observed somewhere, this function is normalized such that
\begin{equation}
\iint \! {\mathcal S}(E_{\mtt{o}}, \eta_{\mtt{o}} \vert E, \eta) \dd{E}_{\mtt{o}} \dd{\eta}_{\mtt{o}} = 1\,.
\end{equation}

\begin{figure*}[h!]
\centering
\begin{subfigure}[b]{0.50\textwidth}
\includegraphics[width=\textwidth]{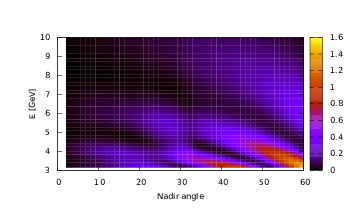}
\caption{}
\label{fig:verNO}
\end{subfigure}\begin{subfigure}[b]{0.50\textwidth}
\includegraphics[width=\textwidth]{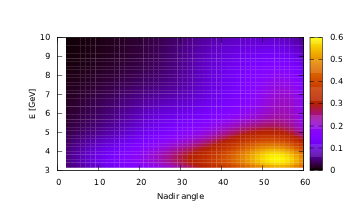}
\caption{}
\label{fig:obsNO}	
\end{subfigure}
\caption{Probability distribution functions of (a) true events and (b) observed events, as functions of energy and nadir angle, for the {\it standard Earth} and  normal ordering of neutrino masses. The same number of bins was used to make both figures, but a much small number of bins was used to calculate the observed events.}
\label{fig:trueobsNO}
\end{figure*}

\begin{figure*}[h!]
\centering
\begin{subfigure}[b]{0.50\textwidth}
\includegraphics[width=\textwidth]{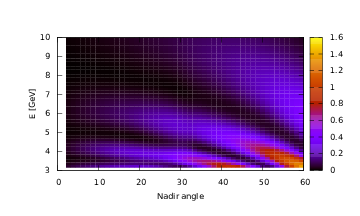}	
\caption{}
\label{fig:verIO}
\end{subfigure}\begin{subfigure}[b]{0.50\textwidth}
\includegraphics[width=\textwidth]{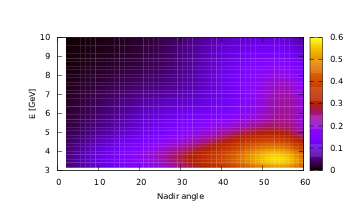}
\caption{}
\label{fig:obsIO}	
\end{subfigure}
\caption{Probability distribution functions of (a) true events and (b) observed events, as functions of energy and nadir angle, for the {\it standard Earth}  and  inverted ordering of neutrino masses. The same number of bins was used to make both figures, but a much small number of bins was used to calculate the observed events.}
\label{fig:trueobsIO}
\end{figure*}

In terms of the resolution function,  the number of observed events $\mathcal{O}^{\,i_{exp}}_{m,n}$ in the bin $(m,n)$ for the $i_{exp}$-th experiment is given by
\begin{equation}
\label{observed}
\mathcal{O}^{\,i_{exp}}_{m,n} = {\mathcal N}_{tot}\sum\limits_{i,j} \iint_{\text{bin} \; m,n} \! \dd{E}_{\mtt{o}} \, \dd{\eta}_{\mtt{o}} \; \iint_{\text{bin} \; i,j} \! \dd{E} \, \dd{\eta}\;
\mathcal{S}(E_{\mtt o}, \eta_{\mtt{o}} \vert E, \eta) f^{\,i_{exp}}_t(E, \eta)\,.
\end{equation}
As an illustration, in Fig. \ref{fig:trueobsNO} and Fig. \ref{fig:trueobsIO} we show the pdf of the true and observed events as functions of the energy and nadir angle, for NO and IO, respectively, in the case of a {\it standard Earth}. The figures were made using the same (large) number of bins for both kind of events, but these numbers generally differ. In what follows, to test how well the hypothesis of different Earth compositions is in agreement with the standard Earth, we take the observed events to be distributed into five angular bins and nine energy bins. 
From Eq. \ref{observed}, for each bin $(m,n)$, with  $m = 1, \cdots, 9$ and $n = 1, \cdots, 5$, we determine the number of events $\mathcal{O}^{\hspace{0.01 cm}{\mtt s}i_{exp}}_{\alpha}$ for a standard Earth and  $\mathcal{O}^{{\hspace{0.02 cm}i_{exp}}}_{\alpha}\!({\rm H}, \varkappa)$ for an alternative Earth with different composition and density. Here $\alpha = 1, \cdots , 45$ label the two-dimensional bins ($m,n$). Thus, for each of these bins we have a sample of the observed events and determinate the corresponding mean values
\begin{eqnarray}
\mathcal {\bar O}^{\mtt s}_{\alpha}&=&\frac1{n_{exp}}\sum_{i_{exp}}\mathcal{O}^{\hspace{0.01 cm}{\mtt s}i_{exp}}_{\alpha}\,, \nonumber\\
\mathcal{\bar O}_{\alpha}({\rm H}, \varkappa)&=&\frac1{n_{exp}}\sum_{i_{exp}}\mathcal{O}^{{\hspace{0.02 cm}i_{exp}}}_{\alpha}\!({\rm H}, \varkappa)\,.
\end{eqnarray}

For Poisson distributed events, in terms of the likelihood function $L$ we construct the negative log-likelihood ratio function as
\begin{eqnarray}
\chi^2_{\lambda} = -\ln\left[\frac{L\left(\mathcal{\bar O}^{\mtt s}_{\alpha};\mathcal{\bar O}_{\alpha}({\rm H}, \varkappa)\right)}{L\left(\mathcal{\bar O}^{\mtt s}_{\alpha},\mathcal{\bar O}^{\mtt s}_{\alpha}\right)}\right]=
2\sum_{\alpha} \left[\mathcal{\bar O}_{\alpha}({\rm H} , \varkappa)-\mathcal{\bar O}^{\mtt s}_{\alpha} \ln\left(\frac{\mathcal{\bar O}^{\mtt s}_{\alpha}}{\mathcal{\bar O}_{\alpha}({\rm H} , \varkappa)}\right)\right]\,.
\end{eqnarray}
According to Wilks' theorem \cite{wilks1938} the $\chi^2_{\lambda}$ distribution can be approximated  by the $\chi^2$ distribution and, from it, the goodness of the fit can be established. The statistical significance of the $\chi^{2}$ test is
given, as usual, by the $\mathbf p$-value: 
\begin{equation}
{\mathbf p}=\int_{\chi^2}^\infty \! f_{\chi}(w,n_{_{\!\rm dof}}) \dd{w}\,,
\end{equation}
where $n_{_{\!\rm dof}}$ is the number of degrees of freedom and
$ f_{\chi}(w,n_{_{\!\rm dof}})$
is the chi-square distribution. In our case, $n_{_{\!\rm dof}} = 9\times5 -2 = 43$.
In this way, we can examine the levels of discrepancy between the standard Earth and different hypothesis about the composition and density of the outer core and the lower mantle. This is done in the next section, where the results and final comments are presented.

\section{Results and Final Comments}
\label{sec:results}

Our goal is to determine to what extent the presence of light elements, in particular hydrogen, in the outer core and mantle can produce measurable effects due to the modifications it introduces in the flavor transformations of atmospheric neutrinos. We also pay attention to how uncertainties in the densities of the deepest regions of the planet can complicate obtaining compositional information. As discussed in Section 2, we rescale the mantle and outer core densities so that the constraints imposed by the Earth's total mass and moment of inertia are satisfied. At the same time, we allow for some changes in the compositions of the outer core and/or the lower mantle $M_1$.

\begin{figure}[h!]
\begin{subfigure}[b]{0.5\textwidth}
\includegraphics[width=\textwidth]{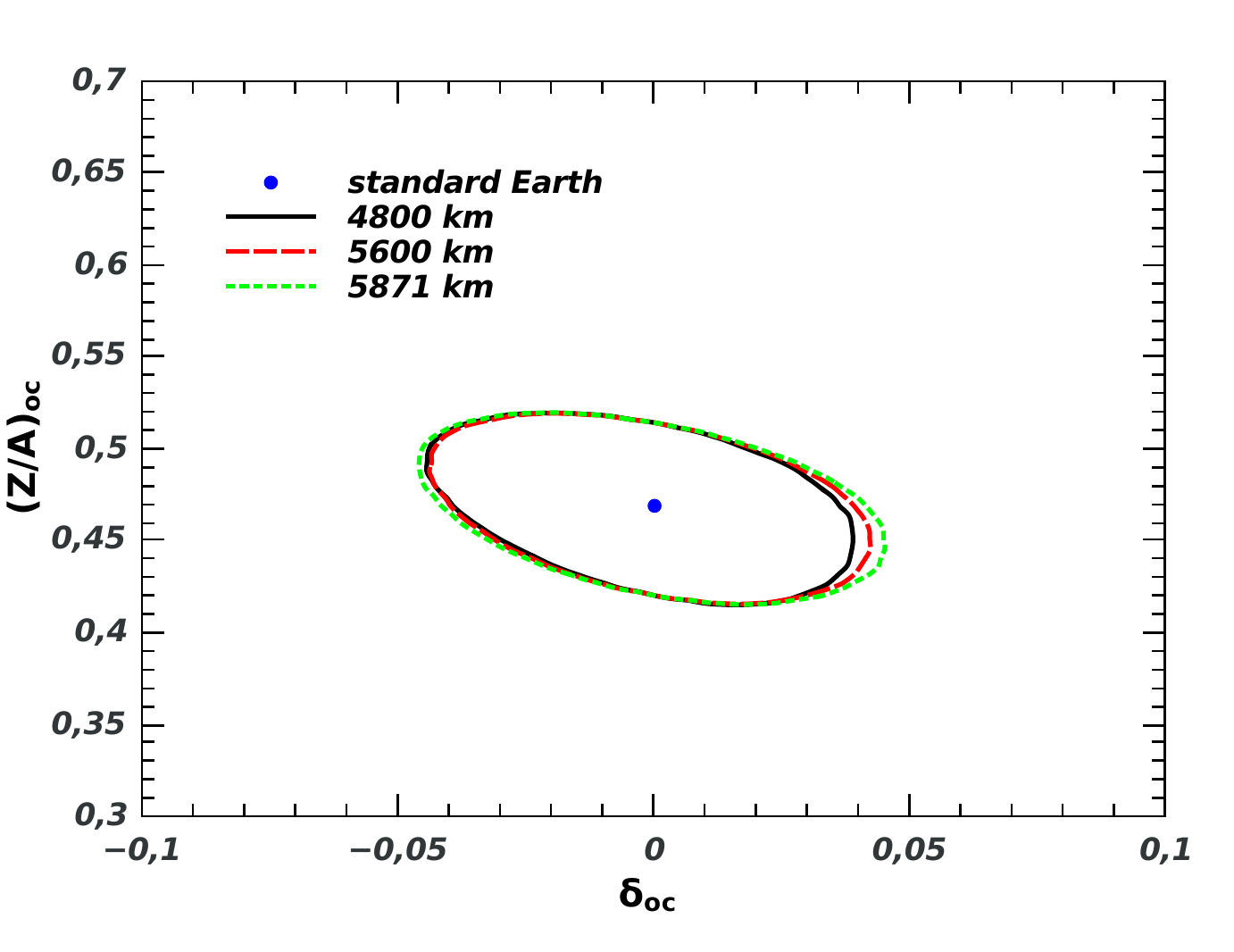}
\caption{Composition and density of the outer-core. }	
\label{fig:c_oc_d_oc_NO}	
\end{subfigure}
\hfill
\begin{subfigure}[b]{0.5\textwidth}
\includegraphics[width=\textwidth]{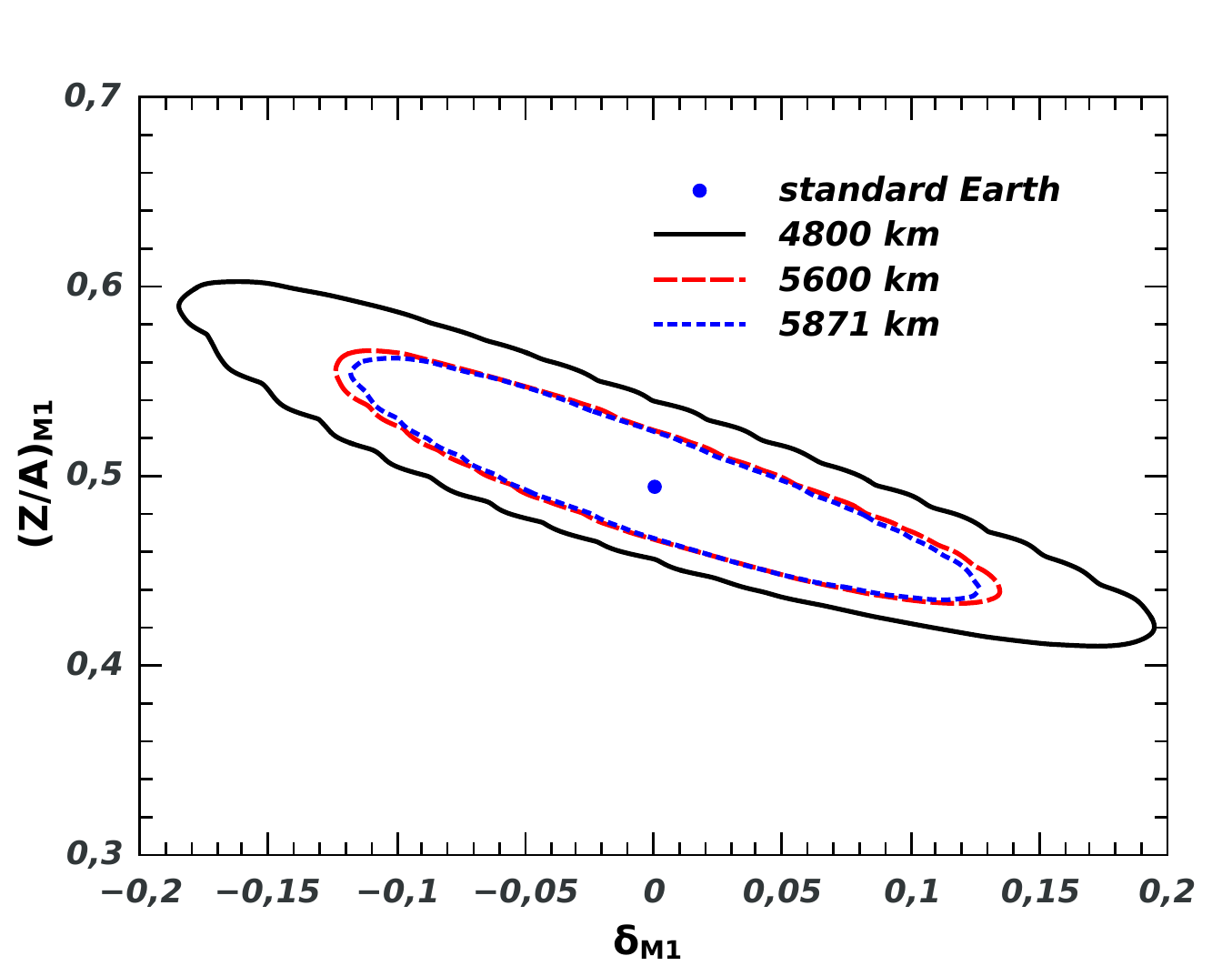}
\caption{Composition and density of the $M_1$ layer. }	
\label{fig:c_M1_d_M1_NO}	
\end{subfigure}
\vskip\baselineskip
\centering
\begin{subfigure}[b]{0.5\textwidth}
\includegraphics[width=\textwidth]{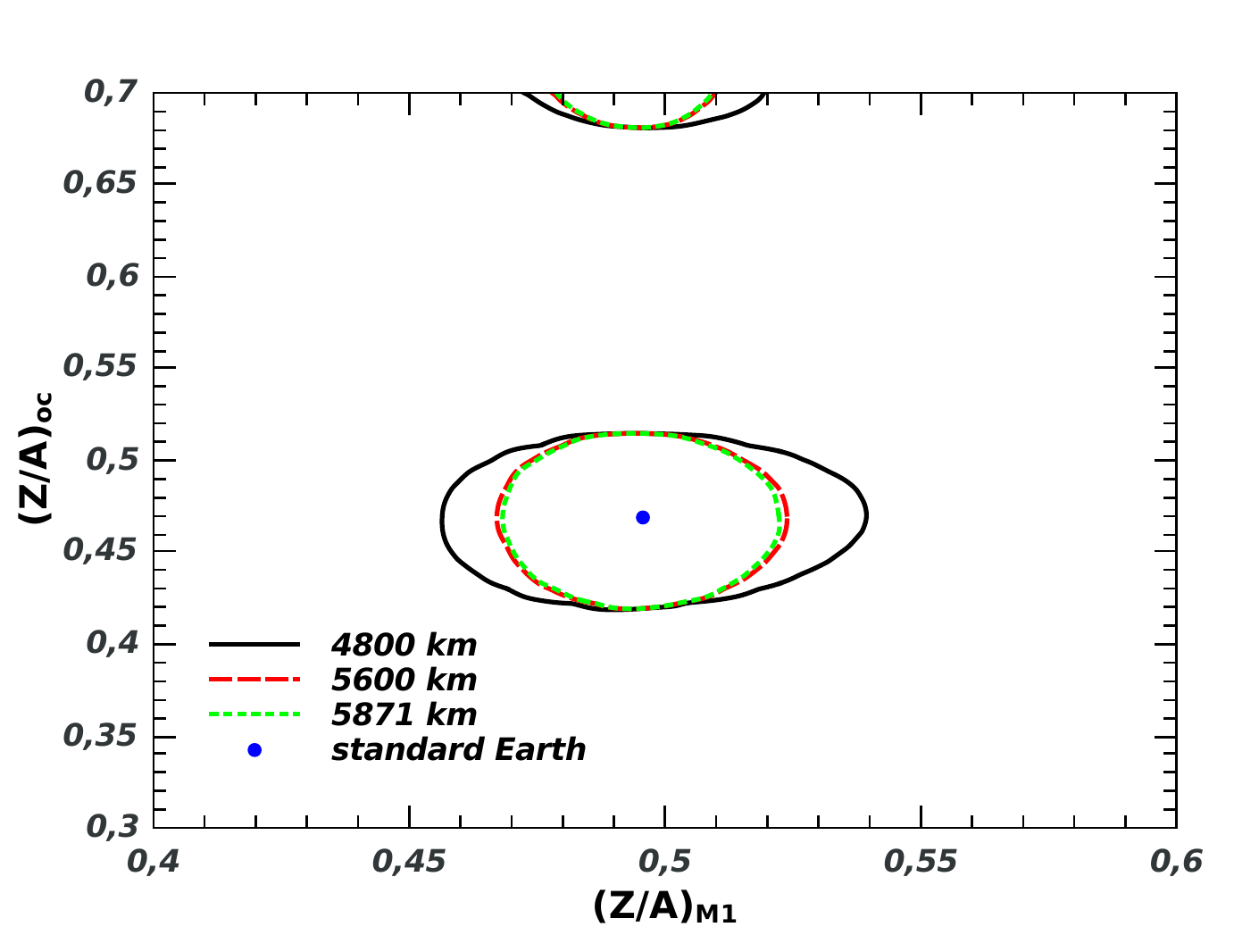}
\caption{Outer-core and $M_1$ compositions.}	
%\label{}	
\end{subfigure}
\caption{Expected 1$\sigma$ regions for combined composition and density measures in the outer-core and  the $M_1$ region, for normal ordering and three values of the $M_1$ radii.}
\label{fig:compo_dens_NO}
\end{figure}

\begin{figure}[h!]
\begin{subfigure}[b]{0.5\textwidth}
\includegraphics[width=\textwidth]{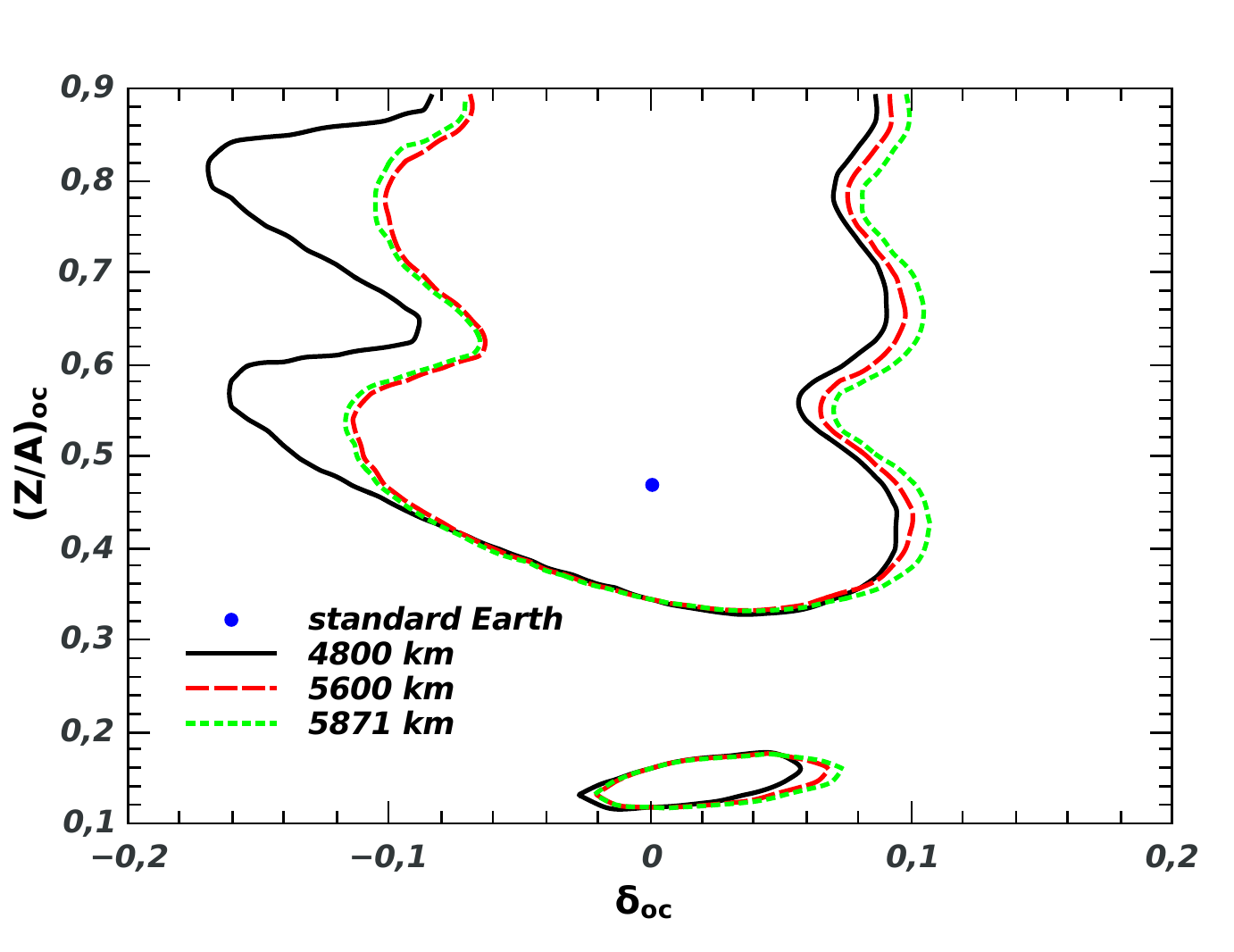}
\caption{Composition and density of the outer-core. }	
\label{fig:c_oc_d_oc_IO}	
\end{subfigure}
\hfill
\begin{subfigure}[b]{0.5\textwidth}
\includegraphics[width=\textwidth]{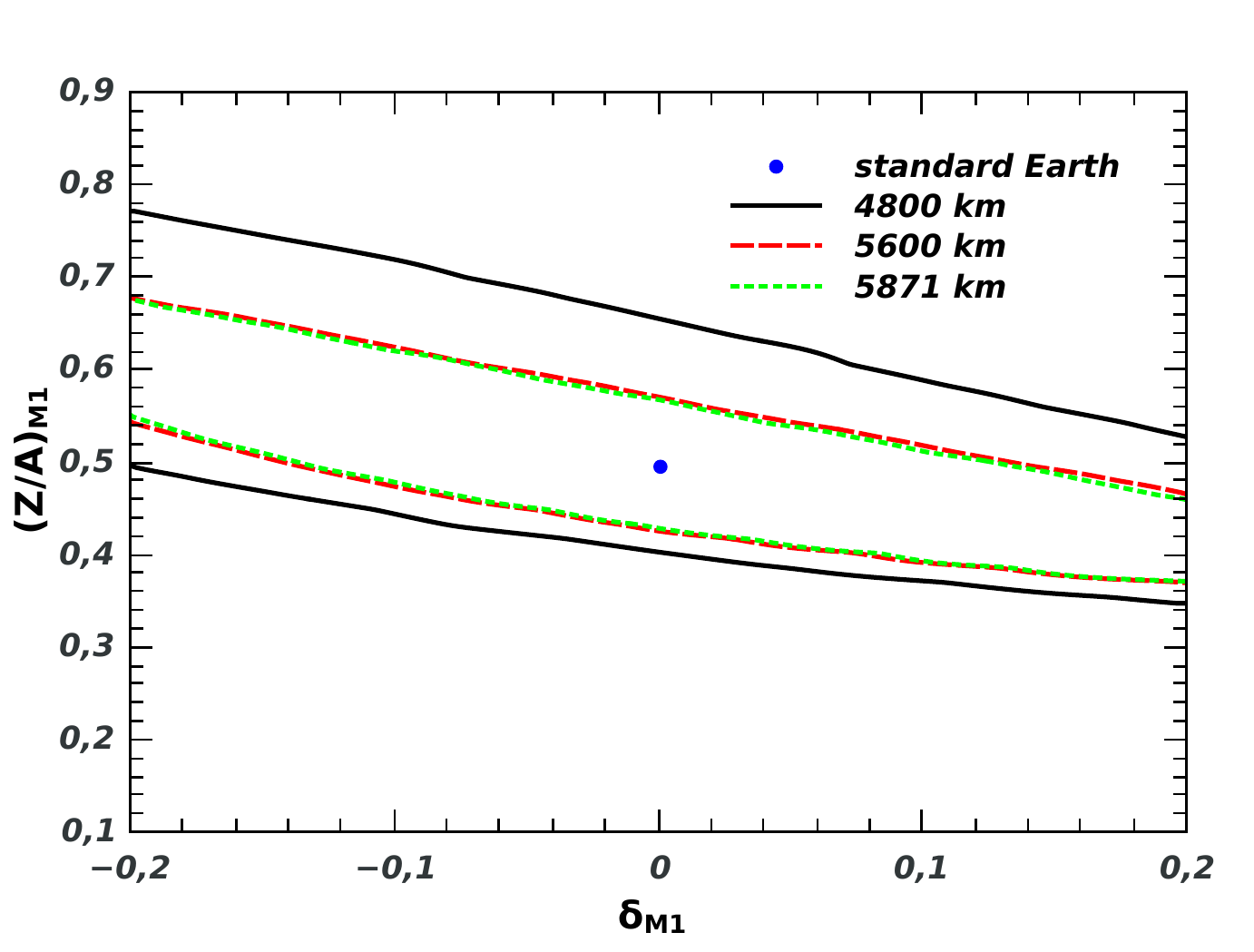}
\caption{Composition and density of the $M_1$ layer. }	
\label{fig:c_M1_d_M1_IO}	
\end{subfigure}
\vskip\baselineskip
\centering
\begin{subfigure}[b]{0.5\textwidth}
\includegraphics[width=\textwidth]{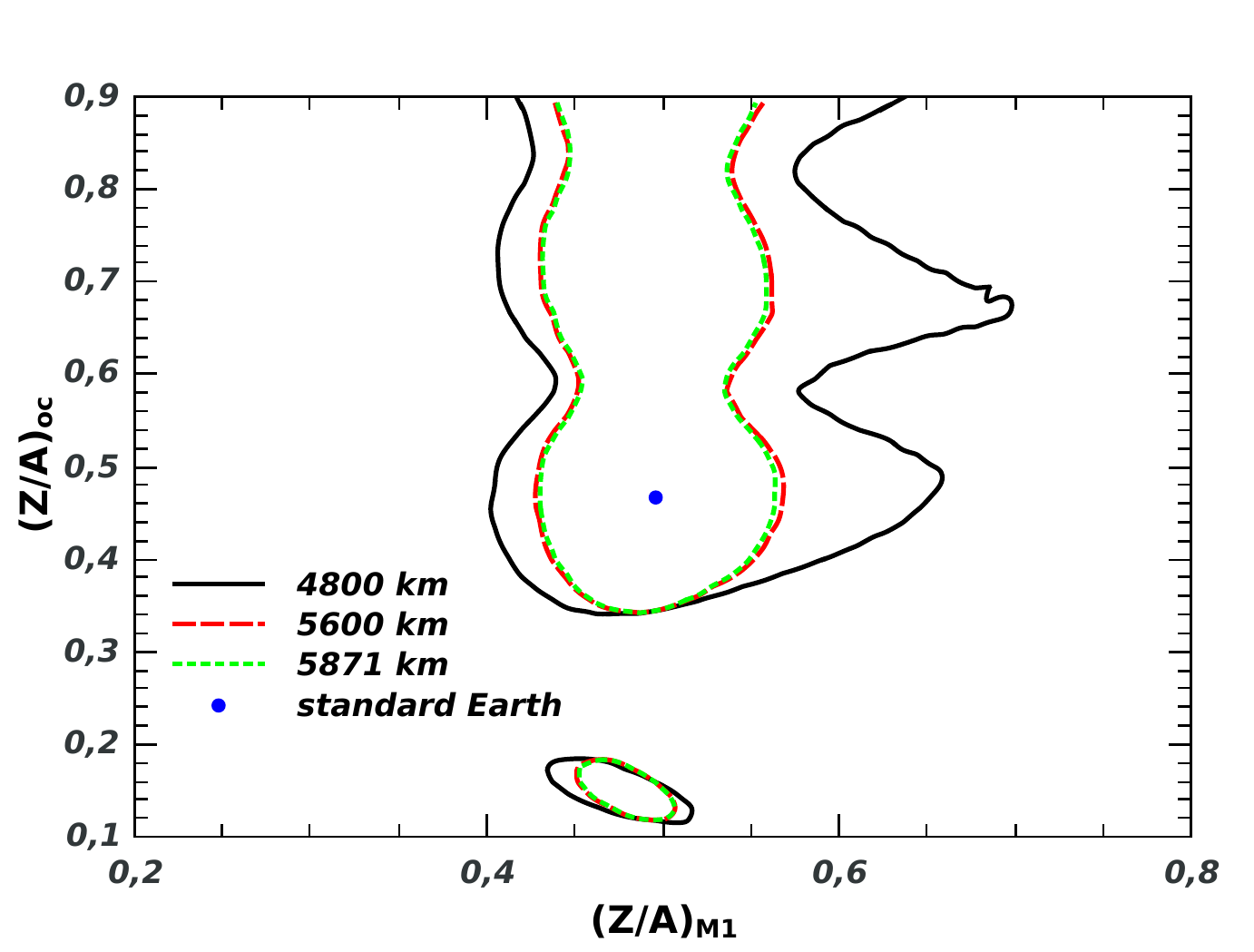}
\caption{Outer-core and $M_1$ composition.}	
%\label{}	
\end{subfigure}
\hfill
\caption{Expected 1$\sigma$ regions for combined composition and density measures in the outer-core and the $M_1$ region, for inverted ordering and three values of the $M_1$ radii.}
\label{fig:compo_dens_IO}
\end{figure}

As the quantities to be fitted, we take the compositions of the outer-core and the lower mantle $M_1$ and the densities of the outer core, $M_1$ and the rest of the mantle $M_2$. To evaluate the effects that changes of two of these quantities have on our observable, we construct regions in several planes: 
($\delta_{\chic{OC}},(Z/A)_{\chic{OC}}$), ($\delta_{M_1},(Z/A)_{M_1}$), and ($(Z/A)_{M_1},(Z/A)_{\chic{OC}}$),  where the statistical significance, given by the ${\mathbf p}$-value, for the discrepancy between the standard and the modified Earth is less than $1\sigma$. In Fig.~(\ref{fig:compo_dens_NO}) and (\ref{fig:compo_dens_IO}) we show these regions for the oscillation parameters given in Table \ref{tab:constants}, for 10 years operation of an 8 Mton detector as ORCA, with resolution parameters $\alpha_{\eta} =0.25$ and $\alpha_{E} = 0.2$. As can be seen, the regions for NO are more restricted than for IO. We consider two different values for the radius of the $M_1$ layer: 5600 km and 5871 km, which correspond to the lower and upper border of the transition zone between the lower and upper mantle.   

\begin{figure}[h!]
\centering
\includegraphics[width=0.75\textwidth]{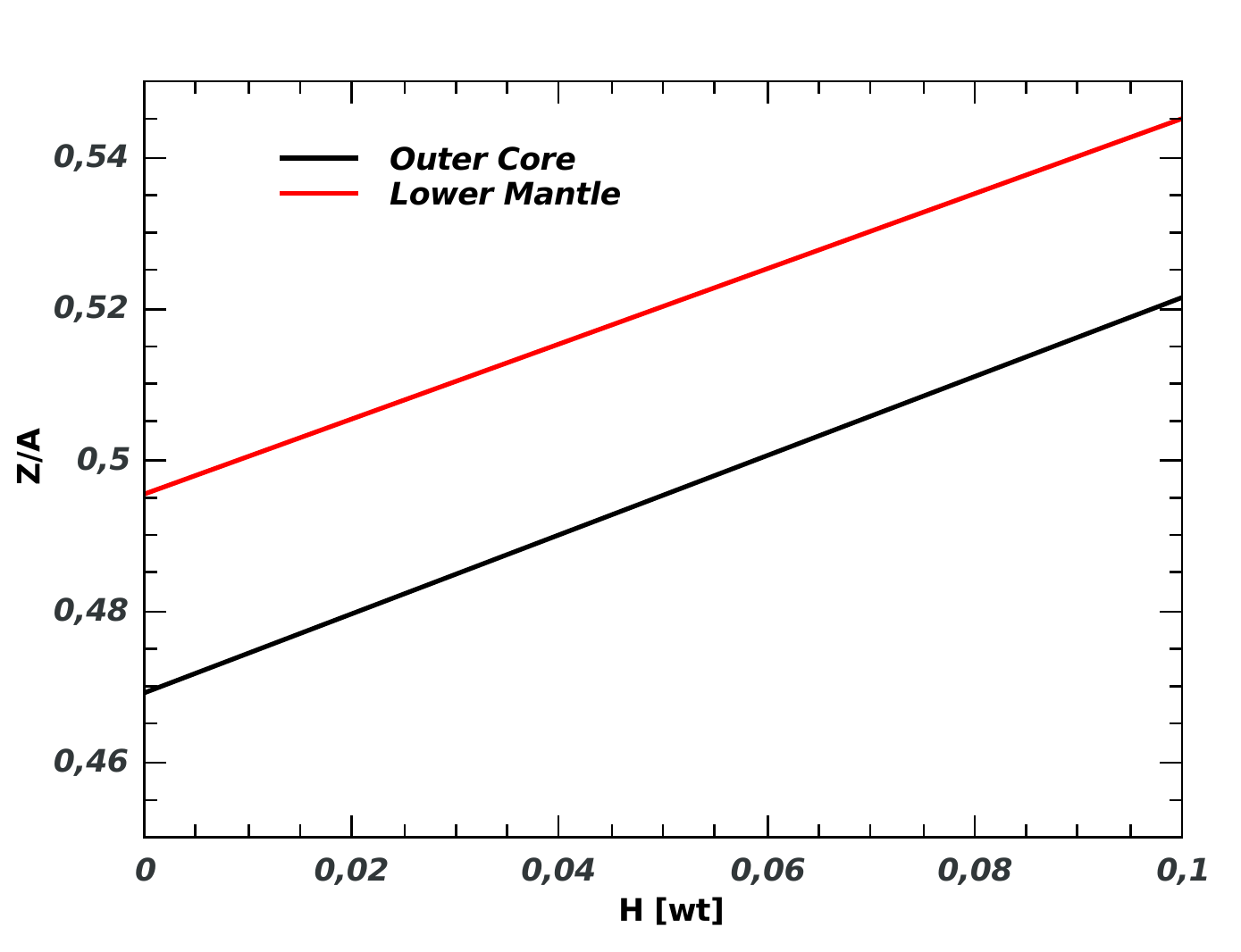}	
%\label{}	
\hfill
\caption{The ratio Z/A as a function of the weight fraction of hydrogen.}
\label{fig:ZA}
\end{figure}

In the graphs of composition versus density, non-zero correlations are observed, indicating that increases in composition are compensated by decreases in density, and vice versa, such that the electron density in the shell does not change. This compensation is not complete since, for the Earth's moment of inertia and mass to remain fixed, variations in density must also occur in other layers traversed by neutrinos and, as a consequence, the allowed regions become closed. The 1$\sigma$ confidence regions for combined measures of Z/A in the outer core and $M_1$  agree with those obtained by \cite{VanElewyck:2017dga}. 
Suppose that the variation of  Z/A in the outer core or mantle is associated only with the abundance of hydrogen, then, as show in Fig.\ref{fig:ZA}, $Z/A=(1-\digamma_{\!H})(Z/A)_0+ \digamma_{\!H}(Z/A)_H$, where $(Z/A)_0$ is the value with no hydrogen and $\digamma_{\!H}$ is the fractional contribution that hydrogen makes to the total mass of the layer. In this case, the 1$\sigma$ regions are compatible with 5-8 wt\% hydrogen in the lower mantle and outer core, respectively, values too high in light of geophysical estimates. 

\begin{figure}[h!]
\centering
\includegraphics[width=0.75\textwidth]{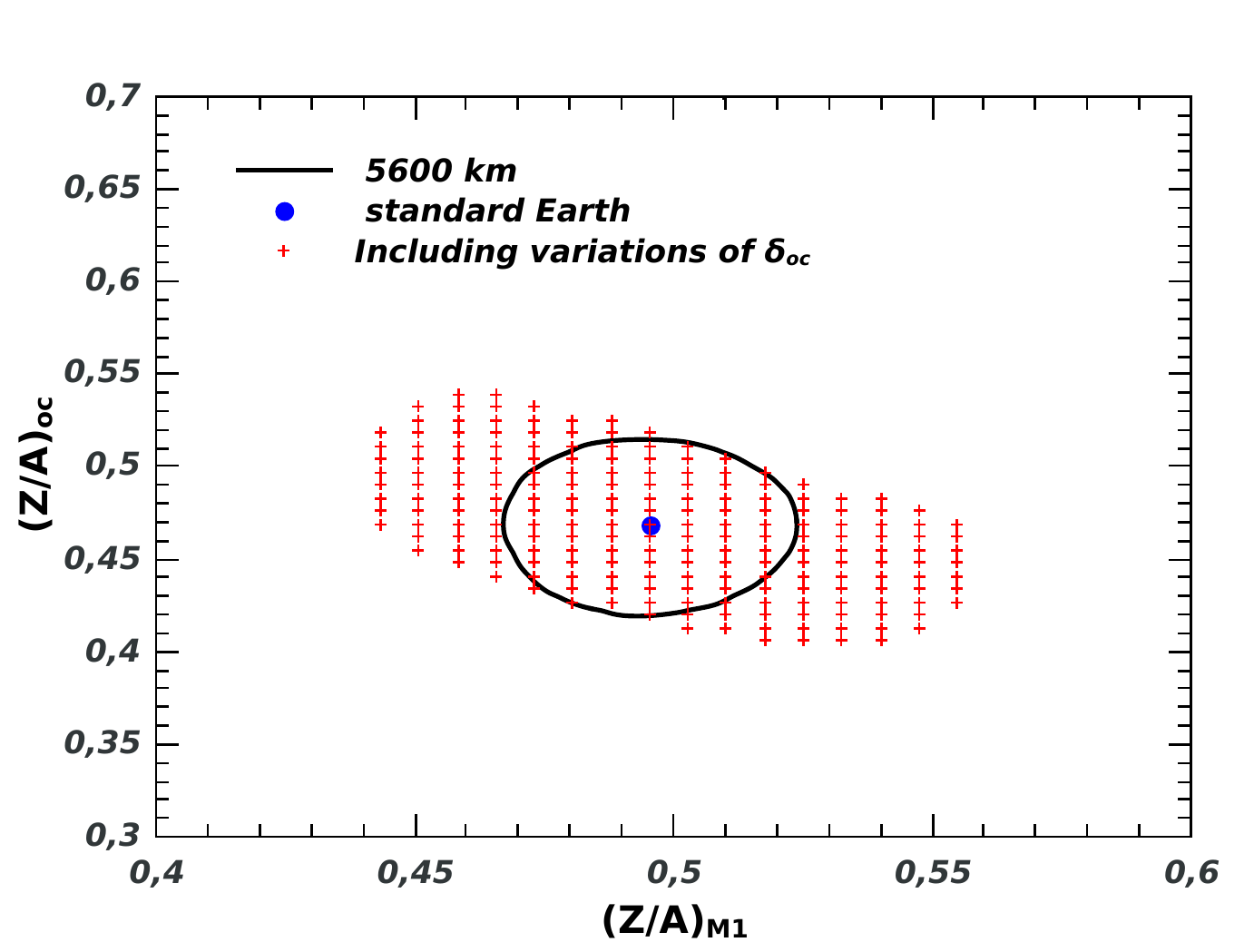}	
%\label{}	
\hfill
\caption{The same region of Fig.\ref{fig:compo_dens_NO}\hs{0.03 cm}c, but including the uncertainties in the outer-core densities.
}
\label{fig:merge}
\end{figure}

As a general rule, for the non fitted parameters we have kept their PREM values. However, for the sake of completeness, we paid some attention to effects from uncertainties in the densities when fitting the compositions. Thus, we have jointly fitted $(Z/A)_{oc}, (Z/A)_{M_1}$ and $\delta_{oc}$ for the outer core density. (Notice, that the densities of the layers $M_1$ and $M_2$ vary in order to maintain the values of the Earth's mass and moment of inertia, as discussed in Section \ref{sec:structure}). The 1$\sigma$ volume obtained in this way is projected onto the composition-composition plane, giving rise to the shaded area in Fig. \ref{fig:merge}. The observed negative correlation is caused by the compensation effect in matter neutrino oscillations, with an increase in composition counterbalanced by a decrease in density, and vice versa, to keep the electron number density unchanged. In the case of the composition-density plots (Fig.\ref{fig:c_oc_d_oc_NO}, Fig.\ref{fig:c_M1_d_M1_NO}, Fig.\ref{fig:c_oc_d_oc_IO}, and Fig.\ref{fig:c_M1_d_M1_IO}), the uncertainty in the outer core density is incorporated in the lower mantle density, since they are related by Eq. \eqref{reldeltaocdeltaM12}.

\begin{figure}[h!]
\centering
\includegraphics[width=0.75\textwidth]{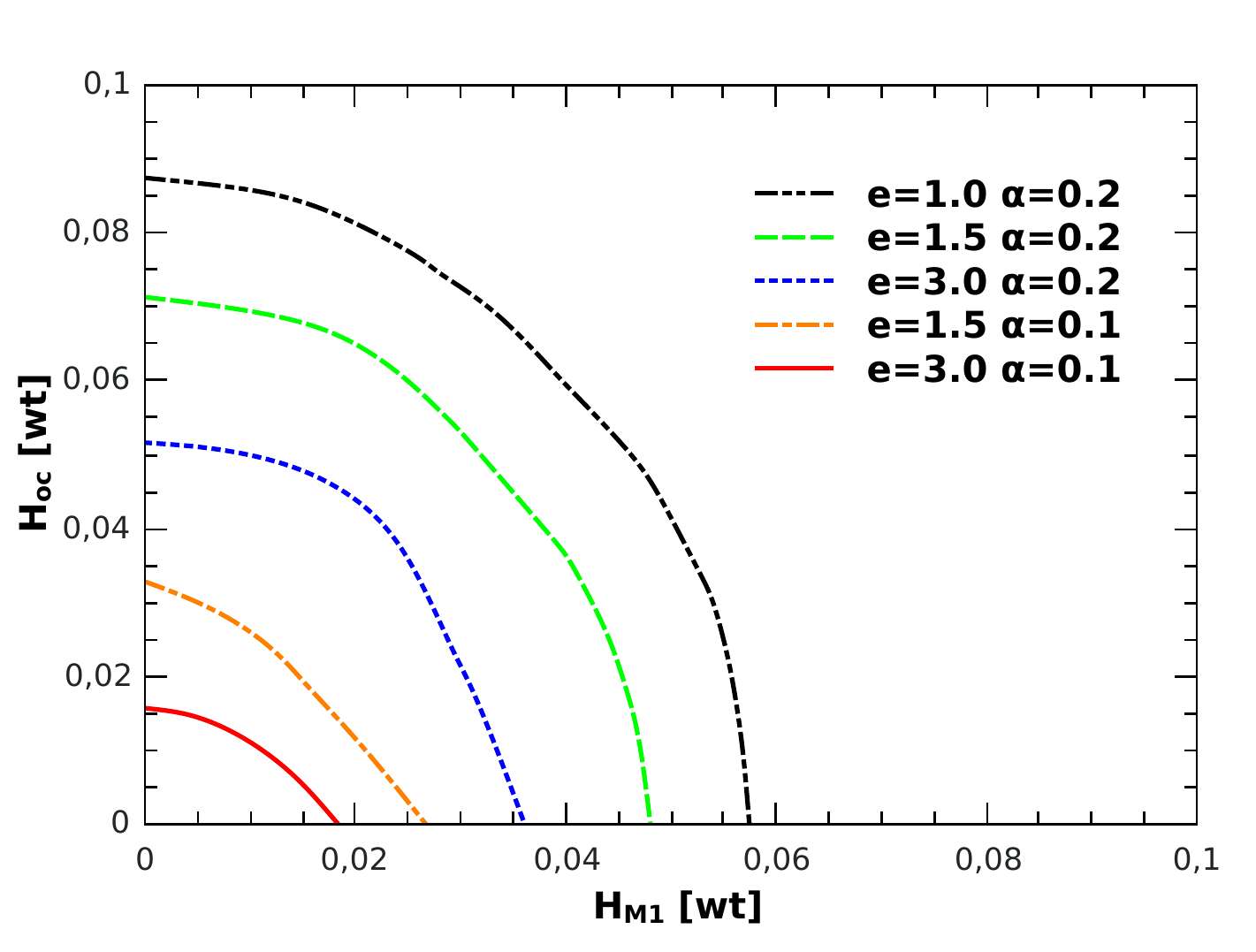}
\caption{Fraction of hidrogen in the outer-core and the lower mantle for extra exposition $\mtt e$ and resolution $\alpha$. }	
\label{fig:resolution}	
\end{figure}

As is evident from Figs. \ref{fig:compo_dens_NO} and \ref{fig:compo_dens_IO}, an experiment like ORCA has a limited potential to reveal a non-standard composition and/or density of the Earth. This worsens at 2 and 3$\sigma$ and for the inverted ordering. A question then arises: how much exposure and how much resolution are required to constraint a non-standard composition? As a partial answer, in Fig.\ref{fig:resolution} we shown the1$\sigma$ regions for different exposures (in unity of ten years of ORCA operation) and angular and energy resolutions equal to $\alpha$. We see that with an exposition of thirty years and resolutions of $0.1$ it is possible to constraint the hydrogen content to 1\%.

\begin{figure}[h!]
\centering
\begin{subfigure}[b]{0.75\textwidth}
\includegraphics[width=\textwidth]{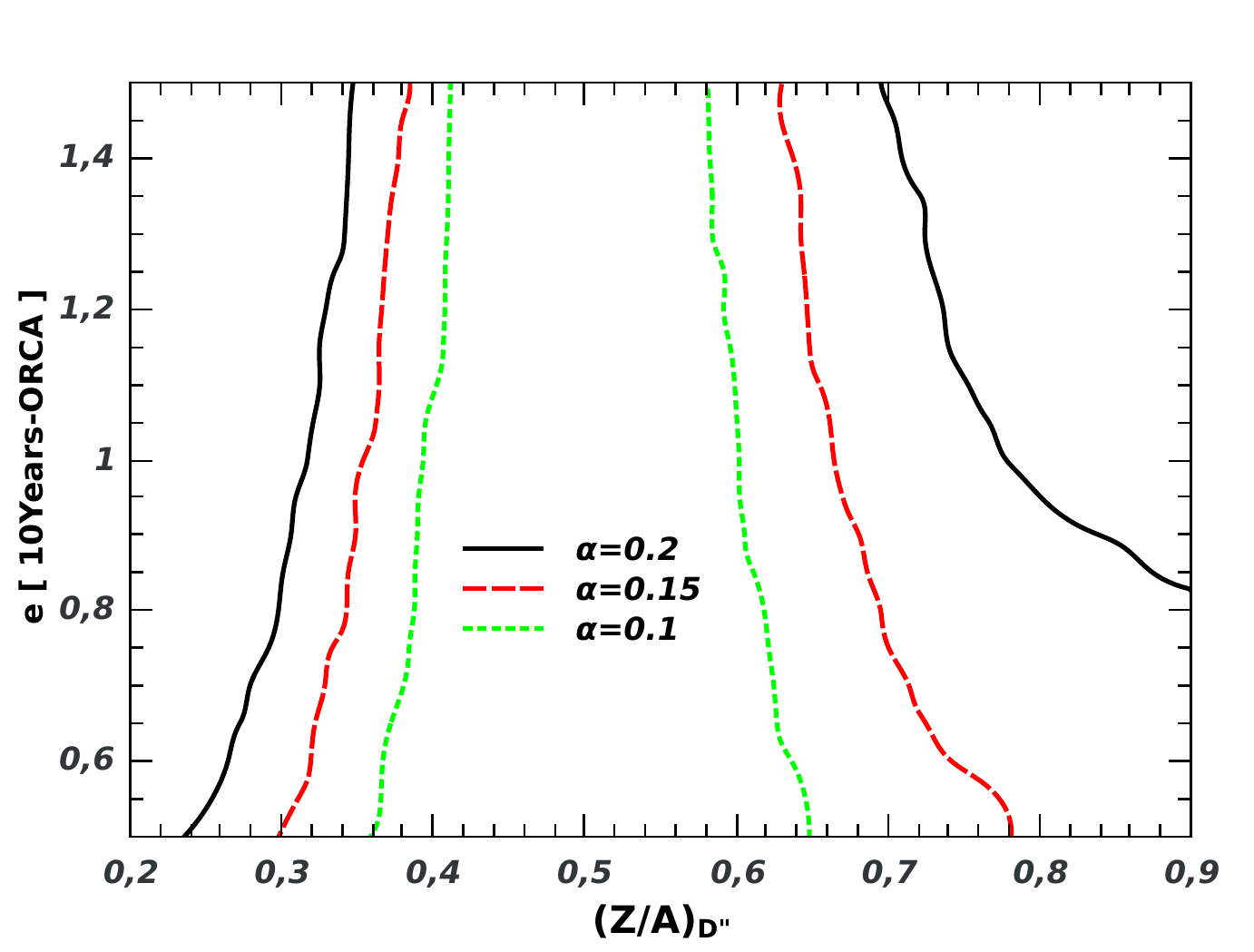}	
\end{subfigure}
%\label{}	
\hfill
\begin{subfigure}[b]{0.75\textwidth}
\includegraphics[width=\textwidth]{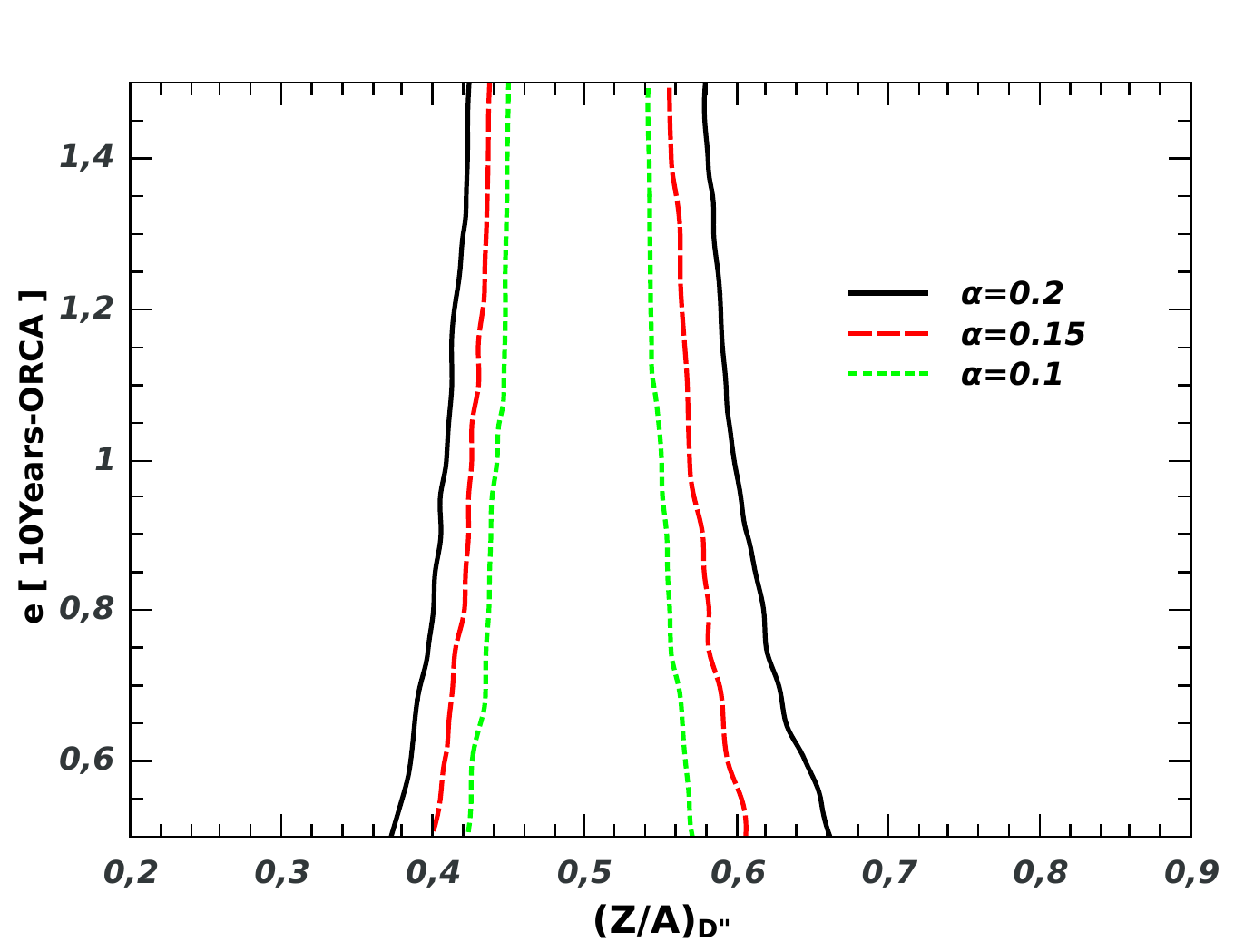}	
\end{subfigure}
\caption{$D^{"}$ composition for different resolutions and size detector. $D^{"}$ thickness $500$ km.}
\label{fig:d2comp}
\end{figure}

Finally, we pay special attention to the $D^{\prime \prime}$ region. Since the thickness of this remote interface between the rocky mantle and the iron core is relatively thin, changes in density and composition have little effect on our observable. Fig.\ref{fig:d2comp} shows the $1\sigma$ regions for the detector size versus the composition $D^{\prime \prime}$, for a thickness of 200 km and various values of the detector resolution. As can be seen, the allowed regions are not very restrictive for reasonable values of resolution and size of the detector. The ability to constrain the composition is significantly improved in the case of a slightly thicker layer (500 km).

In summary, we have studied the possibility of conducting an oscillation tomography of the Earth based on the matter effects on the flavor oscillations of atmospheric neutrinos  propagating through the depths of the planet. Using the $\mu$-like events in a generic large Cherenkov detector as physical observables and making a Monte Carlo simulation of the energy and azimuthal angle distribution of these event, we tested possible variants with respect to a reference geophysical model with the densities as given by PREM and different composition in the outer-core and lower mantle. Unlike previous studies, the procedure we followed in this work allowed us to simultaneously vary the composition and density. When one of these quantities was fixed in the value of the reference geophysical model, our results, shown in Figs. \ref{fig:compo_dens_NO}, are compatible with those obtained by others authors in an uncorrelated way \cite{VanElewyck:2017dga,Borriello:2009ad}. As shown, the study of questions of the type examined here would benefit greatly from the application of phenomena normally studied in the realm of particle physics and the advent of new and improved neutrino telescopes, such as KM3NeT, ORCA, PINGO and HyperK \cite{Adrian-Martinez:2016fdl,Fermani:2019vac,Hyper-Kamiokande:2018ofw,Aarsten:2017}. 

\begin{acknowledgements}
This work was supported in part by DGAPA-UNAM Grant No. PAPIIT IN106322 and by the Consejo Nacional de Investigaciones Cientificas y Tecnicas (CONICET) and the UNMDP Mar del Plata, Argentina.
\end{acknowledgements}

\newpage
\bibliographystyle{spphys}
\bibliography{Tierra.bib}

\begin{thebibliography}{10}
\providecommand{\url}[1]{{#1}}
\providecommand{\urlprefix}{URL }
\expandafter\ifx\csname urlstyle\endcsname\relax
  \providecommand{\doi}[1]{DOI \discretionary{}{}{}#1}\else
  \providecommand{\doi}{DOI \discretionary{}{}{}\begingroup
  \urlstyle{rm}\Url}\fi

\bibitem{tarbuck2009earth}
E.~Tarbuck, F.~Lutgens, D.~Tasa, \emph{{Earth Science}} (Pearson Prentice Hall,
  2009)

\bibitem{fowler:2005}
C.~Fowler, \emph{{The Solid Earth: An Introduction to Global Geophysics (2
  ed.)}} (Cambridge University Pres, 2005)

\bibitem{Lay:1995}
T.~Lay, T.~Wallace, \emph{Modern Global Seismology}, \emph{International
  Geophysics Series}, vol.~58, 1st edn. (Academic Press, San Diego, 1995)

\bibitem{Aki:2002}
K.~Aki, P.G. Richards, \emph{{Quantitative Seismology, 2nd Ed.}} (University
  Science Books, 2002)

\bibitem{Helffrich:2001}
G.R. Helffrich, B.J. Wood, Nature \textbf{412}(6846), 501 (2001).
\newblock \doi{10.1038/35087500}.
\newblock \urlprefix\url{https://doi.org/10.1038/35087500}

\bibitem{Loper:1995}
D.~Loper, L.~Thorne, Journal of Geophysical Research \textbf{100}(B4), 6397
  (1995)

\bibitem{Laywg:1998}
T.~Lay, Q.~Williams, E.~Garnero, Nature \textbf{392}, 461 (1998)

\bibitem{Murakami:2004}
M.~Murakami, K.~Hirose, K.~Kawamura, N.~Sata, Y.~Ohishi, Science
  \textbf{304}(5672), 855 (2004).
\newblock \doi{10.1126/science.1095932}.
\newblock \urlprefix\url{https://doi.org/10.1126%2Fscience.1095932}

\bibitem{Hiroshe:2021}
K.~Hirose, B.~Wood, L.~Vo{\v c}adlo, Nature Reviews Earth \& Environment
  \textbf{2}(9), 645 (2021).
\newblock \doi{10.1038/s43017-021-00203-6}.
\newblock \urlprefix\url{https://doi.org/10.1038/s43017-021-00203-6}

\bibitem{Birch:1964}
F.~Birch, Journal of Geophysical Research \textbf{57}, 4377 (1964)

\bibitem{Poirier:1994}
J.P. Poirier, Physics of the Earth and Planetary Interiors \textbf{85}(3), 319
  (1994).
\newblock \doi{https://doi.org/10.1016/0031-9201(94)90120-1}.
\newblock
  \urlprefix\url{https://www.sciencedirect.com/science/article/pii/0031920194901201}

\bibitem{Litasov:2016}
K.~Litasov, A.~Shatskiy, Russian Geology and Geophysics \textbf{57}, 22 (2016).
\newblock \doi{10.1016/j.rgg.2016.01.003}

\bibitem{Wookey:2008}
J.~Wookey, D.P. Dobson, Philosophical Transactions of the Royal Society A:
  Mathematical, Physical and Engineering Sciences \textbf{366}(1885), 4543
  (2008).
\newblock \doi{10.1098/rsta.2008.0184}.
\newblock \urlprefix\url{https://doi.org/10.1098%2Frsta.2008.0184}

\bibitem{Lister:1995}
J.R. Lister, B.A. Buffett, Physics of the Earth and Planetary Interiors
  \textbf{91}(1), 17 (1995).
\newblock \doi{https://doi.org/10.1016/0031-9201(95)03042-U}.
\newblock
  \urlprefix\url{https://www.sciencedirect.com/science/article/pii/003192019503042U}.
\newblock Study of the Earth's Deep Interior

\bibitem{Roberts:2013}
P.H. Roberts, E.M. King, Reports on Progress in Physics \textbf{76}(9), 096801
  (2013).
\newblock \doi{10.1088/0034-4885/76/9/096801}.
\newblock \urlprefix\url{https://doi.org/10.1088/0034-4885/76/9/096801}

\bibitem{Brodholt:2017}
J.~Brodholt, J.~Badro, Geophysical Research Letters \textbf{44}(16), 8303
  (2017).
\newblock \doi{10.1002/2017gl074261}.
\newblock \urlprefix\url{https://doi.org/10.1002%2F2017gl074261}

\bibitem{Bolt:1995}
B.A. Bolt, Quarterly Journal of the Royal Astronomical Society \textbf{32}(4),
  367 (1991)

\bibitem{Geller:2003}
R.J. Geller, T.~Hara, Nuclear Instruments and Methods in Physics Research
  Section A: Accelerators, Spectrometers, Detectors and Associated Equipment
  \textbf{503}(1-2), 187 (2003).
\newblock \doi{10.1016/s0168-9002(03)00670-3}.
\newblock \urlprefix\url{https://doi.org/10.1016%2Fs0168-9002%2803%2900670-3}

\bibitem{Master:2003}
G.~Masters, D.~Gubbins, Physics of the Earth and Planetary Interiors
  \textbf{140}, 159 (2003).
\newblock \doi{10.1016/j.pepi.2003.07.008}

\bibitem{McDonough:1995}
W.~McDonough, S.~s.~Sun, Chemical Geology \textbf{120}(3), 223 (1995).
\newblock \doi{https://doi.org/10.1016/0009-2541(94)00140-4}.
\newblock
  \urlprefix\url{https://www.sciencedirect.com/science/article/pii/0009254194001404}.
\newblock Chemical Evolution of the Mantle

\bibitem{allegre1995chemical}
C.J. All{\`{e}}gre, J.P. Poirier, E.~Humler, A.W. Hofmann, Earth and Planetary
  Science Letters \textbf{134}(3-4), 515 (1995).
\newblock \doi{10.1016/0012-821x(95)00123-t}.
\newblock \urlprefix\url{https://doi.org/10.1016%2F0012-821x%2895%2900123-t}

\bibitem{Zhang:2016}
Y.~Zhang, T.~Sekine, H.~He, Y.~Yu, F.~Liu, M.~Zhang, Scientific Reports
  \textbf{6}(1) (2016).
\newblock \doi{10.1038/srep22473}.
\newblock \urlprefix\url{https://doi.org/10.1038%2Fsrep22473}

\bibitem{Li:2021}
D.~Alderton, S.A. elias (eds.), \emph{Encyclopedia of Geology}, second edition
  edn.
\newblock 150--163 (Academic Press, 2021)

\bibitem{Hirschmann:2006}
M.M. Hirschmann, Annual Review of Earth and Planetary Sciences \textbf{34}(1),
  629 (2006).
\newblock \doi{10.1146/annurev.earth.34.031405.125211}.
\newblock
  \urlprefix\url{https://doi.org/10.1146/annurev.earth.34.031405.125211}

\bibitem{Peslier:2017}
A.H. Peslier, M.~Sch{\"o}nb{\"a}chler, H.~Busemann, S.I. Karato, Space Science
  Reviews \textbf{212}(1-2), 743 (2017).
\newblock \doi{10.1007/s11214-017-0387-z}.
\newblock \urlprefix\url{https://doi.org/10.1007%2Fs11214-017-0387-z}

\bibitem{Othani:2020}
E.~Ohtani, National Science Review \textbf{7}(1), 224 (2019).
\newblock \doi{10.1093/nsr/nwz071}.
\newblock \urlprefix\url{https://doi.org/10.1093%2Fnsr%2Fnwz071}

\bibitem{Bodnar:2013}
R.~Bodnar, T.~Azbej, S.~Becker, C.~Cannatelli, A.~Fall, M.~Severs, Special
  Paper of the Geological Society of America \textbf{500}, 431 (2013).
\newblock \doi{10.1130/2013.2500(13)}

\bibitem{Garnero:2016}
E.J. Garnero, A.K. McNamara, S.H. Shim, Nature Geoscience \textbf{9}(7), 481
  (2016).
\newblock \doi{10.1038/ngeo2733}.
\newblock \urlprefix\url{https://doi.org/10.1038%2Fngeo2733}

\bibitem{Townsend:2016}
J.P. Townsend, J.~Tsuchiya, C.R. Bina, S.D. Jacobsen, Earth and Planetary
  Science Letters \textbf{454}, 20 (2016).
\newblock \doi{10.1016/j.epsl.2016.08.009}.
\newblock \urlprefix\url{https://doi.org/10.1016%2Fj.epsl.2016.08.009}

\bibitem{Pearson:2014}
D.G. Pearson, F.E. Brenker, F.~Nestola, J.~McNeill, L.~Nasdala, M.T. Hutchison,
  S.~Matveev, K.~Mather, G.~Silversmit, S.~Schmitz, B.~Vekemans, L.~Vincze,
  Nature \textbf{507}(7491), 221 (2014).
\newblock \doi{10.1038/nature13080}.
\newblock \urlprefix\url{https://doi.org/10.1038%2Fnature13080}

\bibitem{tschauner2018ice}
O.~Tschauner, S.~Huang, E.~Greenberg, V.~Prakapenka, C.~Ma, G.~Rossman,
  A.~Shen, D.~Zhang, M.~Newville, A.~Lanzirotti, et~al., Science
  \textbf{359}(6380), 1136 (2018)

\bibitem{karato2011water}
S.~ichiro Karato, Earth and Planetary Science Letters \textbf{301}(3), 413
  (2011).
\newblock \doi{https://doi.org/10.1016/j.epsl.2010.11.038}.
\newblock
  \urlprefix\url{https://www.sciencedirect.com/science/article/pii/S0012821X10007454}

\bibitem{karato:2020}
S.i. Karato, B.~Karki, J.~Park, Progress in Earth and Planetary Science
  \textbf{7}(1), 76 (2020).
\newblock \doi{10.1186/s40645-020-00379-3}.
\newblock \urlprefix\url{https://doi.org/10.1186/s40645-020-00379-3}

\bibitem{Fei:2017}
H.~Fei, D.~Yamazaki, M.~Sakurai, N.~Miyajima, H.~Ohfuji, T.~Katsura,
  T.~Yamamoto, Science Advances \textbf{3}(6) (2017).
\newblock \doi{10.1126/sciadv.1603024}.
\newblock \urlprefix\url{https://doi.org/10.1126%2Fsciadv.1603024}

\bibitem{munch:2020}
F.D. Munch, A.V. Grayver, M.~Guzavina, A.V. Kuvshinov, A.~Khan, Geophysical
  Research Letters \textbf{47}(10), e2020GL087222 (2020)

\bibitem{Winter:2006vg}
W.~Winter, Earth Moon Planets \textbf{99}, 285 (2006).
\newblock \doi{10.1007/s11038-006-9101-y}

\bibitem{Volkova:1974xa}
L.~Volkova, G.~Zatsepin, Izv. Akad. Nauk Ser. Fiz \textbf{38N5}, 1060 (1974)

\bibitem{Wilson:1983an}
T.L. Wilson, Nature \textbf{309}, 38 (1984).
\newblock \doi{10.1038/309038a0}

\bibitem{Ralston:1999fz}
J.P. Ralston, P.~Jain, G.M. Frichter, in \emph{{26th International Cosmic Ray
  Conference}}, vol.~2 (1999), vol.~2, p. 504

\bibitem{Jain:1999kp}
P.~Jain, J.P. Ralston, G.M. Frichter, Astropart. Phys \textbf{12}, 193 (1999).
\newblock \doi{10.1016/S0927-6505(99)00088-2}

\bibitem{GonzalezGarcia:2007gg}
M.~Gonzalez-Garcia, F.~Halzen, M.~Maltoni, H.K. Tanaka, Phys. Rev. Lett
  \textbf{100}, 061802 (2008).
\newblock \doi{10.1103/PhysRevLett.100.061802}

\bibitem{Reynoso:2004dt}
M.M. Reynoso, O.A. Sampayo, Astropart. Phys \textbf{21}, 315 (2004).
\newblock \doi{10.1016/j.astropartphys.2004.01.003}

\bibitem{Romero:2011zzb}
I.~Romero, O.A. Sampayo, Eur. Phys. J. C \textbf{71}, 1696 (2011).
\newblock \doi{10.1140/epjc/s10052-011-1696-0}

\bibitem{Donini:2018tsg}
A.~Donini, S.~Palomares-Ruiz, J.~Salvado, Nature Phys. \textbf{15}(1), 37
  (2019).
\newblock \doi{10.1038/s41567-018-0319-1}

\bibitem{Wolfenstein:1977ue}
L.~Wolfenstein, Phys. Rev. D \textbf{17}, 2369 (1978).
\newblock \doi{10.1103/PhysRevD.17.2369}

\bibitem{Barger:1980tf}
V.D. Barger, K.~Whisnant, S.~Pakvasa, R.~Phillips, Phys. Rev. D \textbf{22},
  2718 (1980).
\newblock \doi{10.1103/PhysRevD.22.2718}

\bibitem{Mikheev:1986gs}
S.~Mikheyev, A.~Smirnov, Sov. J. Nucl. Phys. \textbf{42}, 913 (1985)

\bibitem{Nicolaidis:1987fe}
A.~Nicolaidis, Phys. Lett. B \textbf{200}, 553 (1988).
\newblock \doi{10.1016/0370-2693(88)90170-0}

\bibitem{Nicolaidis:1990jm}
A.~Nicolaidis, M.~Jannane, A.~Tarantola, J. Geophys. Res. \textbf{96}(B13),
  21811 (1991).
\newblock \doi{10.1029/91JB01835}

\bibitem{Borriello:2009ad}
E.~Borriello, G.~Mangano, A.~Marotta, G.~Miele, P.~Migliozzi, C.~Moura,
  S.~Pastor, O.~Pisanti, P.E. Strolin, JCAP \textbf{06}, 030 (2009).
\newblock \doi{10.1088/1475-7516/2009/06/030}

\bibitem{Winter:2015zwx}
W.~Winter, Nucl. Phys. B \textbf{908}, 250 (2016).
\newblock \doi{10.1016/j.nuclphysb.2016.03.033}

\bibitem{Rott:2015kwa}
C.~Rott, A.~Taketa, D.~Bose, Sci. Rep \textbf{5}, 15225 (2015).
\newblock \doi{10.1038/srep15225}

\bibitem{VanElewyck:2017dga}
S.~Bourret, J.A. Coelho, V.~Van~Elewyck, PoS \textbf{ICRC2017}, 1020 (2018).
\newblock \doi{10.22323/1.301.1020}

\bibitem{Bourret:2017tkw}
S.~Bourret, J.a.A.B. Coelho, V.~Van~Elewyck, J. Phys. Conf. Ser.
  \textbf{888}(1), 012114 (2017).
\newblock \doi{10.1088/1742-6596/888/1/012114}

\bibitem{DOlivo:2020ssf}
J.C. D'Olivo, J.A. Herrera~Lara, I.~Romero, O.A. Sampayo, G.~Zapata, Eur. Phys.
  J. C \textbf{80}(10), 1001 (2020).
\newblock \doi{10.1140/epjc/s10052-020-08585-5}

\bibitem{Kelly:2021jfs}
K.J. Kelly, P.A.N. Machado, I.~Martinez-Soler, Y.F. Perez-Gonzalez,
  FERMILAB-PUB-21-459-T, NUHEP-TH/21-15, arXiv/2110.00003  (2021)

\bibitem{Denton:2021rgt}
P.B. Denton, R.~Pestes, Phys. Rev. D \textbf{104}(11), 113007 (2021).
\newblock \doi{10.1103/PhysRevD.104.113007}

\bibitem{Upadhyay:2021kzf}
A.K. Upadhyay, A.~Kumar, S.K. Agarwalla, A.~Dighe, IP/BBSR/2021-12,
  TIFR/TH/21-22, arXiv/2112.14201  (2021)

\bibitem{dziewonski1981preliminary}
A.M. Dziewonski, D.L. Anderson, Physics of the earth and planetary interiors
  \textbf{25}(4), 297 (1981)

\bibitem{Kennett:1995}
B.L.N. Kennett, E.R. Engdahl, R.~Buland, Geophysical Journal International
  \textbf{122}(1), 108 (1995).
\newblock \doi{10.1111/j.1365-246X.1995.tb03540.x}.
\newblock \urlprefix\url{https://doi.org/10.1111/j.1365-246X.1995.tb03540.x}

\bibitem{Williams:1994}
J.~Williams, The Astronomical Journal \textbf{108}, 711 (1994).
\newblock \doi{10.1086/117108}

\bibitem{Capozzi:2020}
F.~Capozzi, E.~Di~Valentino, E.~Lisi, A.~Marrone, A.~Melchiorri, A.~Palazzo,
  Phys. Rev. D \textbf{101}(11), 116013 (2020).
\newblock \doi{10.1103/PhysRevD.101.116013}.
\newblock \urlprefix\url{https://link.aps.org/doi/10.1103/PhysRevD.101.116013}

\bibitem{Esteban:2020cvm}
I.~Esteban, M.C. Gonzalez-Garcia, M.~Maltoni, T.~Schwetz, A.~Zhou, JHEP
  \textbf{09}, 178 (2020).
\newblock \doi{10.1007/JHEP09(2020)178}

\bibitem{deSalas:2020pgw}
P.F. de~Salas, D.V. Forero, S.~Gariazzo, P.~Mart\'\i{}nez-Mirav\'e, O.~Mena,
  C.A. Ternes, M.~T\'ortola, J.W.F. Valle, JHEP \textbf{02}, 071 (2021).
\newblock \doi{10.1007/JHEP02(2021)071}

\bibitem{Bilenky:1980cx}
S.M. Bilenky, J.~Hosek, S.~Petcov, Phys. Lett. B \textbf{94}, 495 (1980).
\newblock \doi{10.1016/0370-2693(80)90927-2}

\bibitem{Ohlsson:1999xb}
T.~Ohlsson, H.~Snellman, J. Math. Phys. \textbf{41}, 2768 (2000).
\newblock \doi{10.1063/1.533270}.
\newblock [Erratum: J.Math.Phys. 42, 2345 (2001)]

\bibitem{Stavropoulos:2021hir}
D.~Stavropoulos, V.~Pestel, Z.~Aly, E.~Tzamariudaki, C.~Markou, PoS
  \textbf{ICRC2021}, 1125 (2021).
\newblock \doi{10.22323/1.395.1125}

\bibitem{Kalaczynski:2021ytv}
P.~Kalaczy\'nski, PoS \textbf{ICHEP2020}, 149 (2021).
\newblock \doi{10.22323/1.390.0149}

\bibitem{Atmnu:1996}
V.~Agrawal, T.~Gaisser, P.~Lipari, T.~Stanev, Phys. Rev. D \textbf{53}, 1314
  (1996).
\newblock \doi{10.1103/PhysRevD.53.1314}

\bibitem{Formaggio:2013kya}
J.~Formaggio, G.~Zeller, Rev. Mod. Phys. \textbf{84}, 1307 (2012).
\newblock \doi{10.1103/RevModPhys.84.1307}

\bibitem{wilks1938}
S.S. Wilks, Ann. Math. Statist. \textbf{9}(1), 60 (1938).
\newblock \doi{10.1214/aoms/1177732360}.
\newblock \urlprefix\url{https://doi.org/10.1214/aoms/1177732360}

\bibitem{Adrian-Martinez:2016fdl}
S.~Adrian-Martinez, et~al., J. Phys. \textbf{G43}(8), 084001 (2016).
\newblock \doi{10.1088/0954-3899/43/8/084001}

\bibitem{Fermani:2019vac}
P.~Fermani, I.~Di~Palma, EPJ Web Conf \textbf{209}, 01006 (2019).
\newblock \doi{10.1051/epjconf/201920901006}

\bibitem{Hyper-Kamiokande:2018ofw}
K.~Abe, et~al., arxiv/1805.04163  (2018)

\bibitem{Aarsten:2017}
M.~Aartsen, et~al., Journal of Physics G \textbf{44}, 054006 (2017).
\newblock \doi{10.1088/1361-6471/44/5/054006}

\end{thebibliography}
\end{document}